\documentclass[aps,prb,floatfix,amsmath,amssymb,eqsecnum,nofootinbib,superscriptaddress]{revtex4}
\usepackage{graphicx}
\usepackage{dcolumn}
\usepackage{bm}
\newcommand{\beq}{\begin{equation}}
\newcommand{\eeq}{\end{equation}}
\newcommand{\ba}{\begin{array}{ccc}}
\newcommand{\ea}{\end{array}}
\newcommand{\nn}{\nonumber \\}
\newcommand{\bx}{{\bm x}}
\newcommand{\by}{{\bm y}}
\newcommand{\bk}{{\bm k}}
\newcommand{\bp}{{\bm p}}
\newcommand{\bq}{{\bm q}}
\def\bea{\begin{eqnarray}}
\def\eea{\end{eqnarray}}

\usepackage{setspace} 
\usepackage{amsmath}
\usepackage{amssymb}

\begin{document}

\title{Spectral functions of the Higgs mode\\ near two-dimensional quantum critical points}
  
\author{Daniel Podolsky}
\affiliation{Physics Department, Technion, 32000 Haifa, Israel}

\author{Subir Sachdev}
\affiliation{Department of Physics, Harvard University, Cambridge MA
02138, USA}
\affiliation{Instituut-Lorentz for Theoretical Physics, Universiteit Leiden, 
P.O. Box 9506, 2300 RA Leiden, The Netherlands}

\date{\today \\
\vspace{1.6in}}

\begin{abstract}
We study the Higgs excitation in the Goldstone phase of the relativistic O($N$) model in two
spatial dimensions at zero temperature. The response functions of the order parameter, and its magnitude-squared,
become universal functions of frequency 
in the vicinity of the quantum critical point described by the Wilson-Fisher fixed point, and
we compute them to next-to-leading order in $1/N$.
The Higgs particle has an infrared singular decay to gapless Goldstone excitations, and its
response functions are characterized by a pole in the lower half of the complex frequency plane.
The pole acquires a non-zero real part only at  next-to-leading order in $1/N$, demonstrating that the Higgs excitation has
an oscillatory component even in the scaling limit. Both the real and imaginary parts of the pole position vanish
with the correlation length exponent $\nu$ upon approaching the critical point. 
We present evidence that the spectral density of the O($N$)-invariant
amplitude-squared of the order parameter has a peak at a non-zero frequency in the scaling limit. We connect our results to
recent experimental studies of the superfluid-insulator quantum phase transition of ultracold bosonic atoms in optical lattices.
\end{abstract}

\maketitle

\section{Introduction}
\label{sec:intro}

Quantum critical points  described by relativistic field theories have long been the focus of theoretical study in
condensed matter physics.\cite{chn,fwgf,csy,zwerger,senthil,ssrelax,paa}
A number of experimental realizations of such critical points have also appeared in recent years. The critical point of
the quantum Ising model in spatial dimension $d=1$ has been realized in the ferromagnetic insulator\cite{coldea}
CoNb$_2$O$_6$, and also in tilted
lattices\cite{sstilt} of ultracold bosons.\cite{simon} Neutron scattering experiments\cite{ruegg} on pressure-tuned 
TlCuCl$_3$
have mapped the excitations across the quantum critical point associated with the onset of N\'eel order
in a dimerized antiferromagnet in $d=3$. 
The organic insulator (TMTTF)$_2$PF$_6$ has a quantum phase transition \cite{brown} involving onset of N\'eel order under pressure
in $d=2$, but the excitations have not yet been completely studied. Ultracold bosonic atoms in optical lattices provide remarkable
tunable realizations\cite{greiner,spielman} of the superfluid-Mott insulator quantum phase transition in $d=2$ and $d=3$, whose properties can be studied closely using {\em in situ} imaging techniques.\cite{gemelke,sherson,bakr,endresString,endres}

This paper will study the excitation spectrum in the 
vicinity of the quantum critical point of the relativistic O($N$) model in two spatial dimensions.
The critical point itself is a conformal field theory in three spacetime dimensions (a CFT3), associated with the
Wilson-Fisher fixed point.\cite{wf} Extensive studies of the excitation spectrum have appeared in 
previous work,\cite{chn,csy,zwerger,ssrelax,paa} both at $T=0$ and $T>0$. 
Our focus here will be on the nature of the amplitude (`Higgs') excitation 
in the Goldstone phase with broken O($N$) symmetry at $T=0$. 
It is generally believed that the excitation structure of this Goldstone phase is well understood, with $N-1$ gapless Goldstone
modes which behave like free particles at long wavelengths. However, there remain key open questions on the nature of
the amplitude mode. In mean field theory, there is a sharp, gapped excitation in the longitudinal response, which may be identified
with the Higgs particle. Beyond the mean field level, this excitation can emit multiple Goldstone excitations even at $T=0$, and so 
the  stability of the Higgs is not guaranteed. The Higgs has been argued\cite{ssbook,sushkov} to be marginally stable in $d=3$.
In $d=2$, there are infrared singularities\cite{csy,ssrelax,zwerger} in the spin wave emission process, which indicate
that the longitudinal response is overdamped, and imply that
the ultimate nature of the
amplitude mode is a strong-coupling problem. The question of whether a finite frequency peak appears at the Higgs
energy can depend\cite{lindner,paa} upon the specific observable being used to probe the amplitude mode.

Here, we will present a $1/N$ expansion
of the spectrum of the Higgs excitation to next-to-leading order, and find some qualitatively new features. We show
that the Higgs response functions are characterized by a pole in the lower-half complex frequency plane, and this pole
acquires a non-zero real part at order $1/N$. We also find an associated structure in the spectral functions on the real frequency axis.
Our results can serve as tests
of the strong-coupling dynamics of the CFT3 in experiments. For instance, recent experiments on ultracold bosons in optical lattices \cite{endres} have studied the low energy response near the $2D$ Mott-superfluid phase transition, described at integer boson filling by an $O(2)$ critical point with relativistic dynamics in 2+1 dimensions.

We represent the O($N$) order parameter field by the $N$-component real field $\phi_\alpha$, with $\alpha = 1 \ldots N$.
Let us orient the condensate in the Goldstone phase in the $\alpha=N$ direction, so that $\langle \phi_\alpha \rangle
\propto \delta_{\alpha,N}$. 
As emphasized in a recent paper,\cite{paa} it is useful to consider two separate observable probes of the Higgs spectrum of fluctuations
about this condensate.

The first observable concerns the longitudinal fluctuation of the order parameter, represented by 
$\phi_N$, and we denote its connected two-point correlator by the susceptibility $\chi$, where 
\beq
\chi (p) = \int d^3 x \, e^{-i \bp \cdot \bx} \left[ \langle \phi_N (\bx) \phi_N (0) \rangle -  \langle \phi_N (\bx) \rangle \langle \phi_N (0) \rangle\right]
\eeq
where $\bx$ is a spacetime co-ordinate, $\bp$ is spacetime momentum. We will work largely in imaginary time, and represent
$p \equiv |\bp|$, $x \equiv |{\bx}|$, etc. The real time retarded response function, $\chi (\omega)$ is obtained by the analytic continuation
$p \rightarrow - i \omega$ where
$\omega$ is a measurement
frequency and the momentum is zero. (Because of the relativistic invariance of the theory at $T=0$, the susceptibility at non-zero momenta $k$ can be deduced
by $\omega \rightarrow \sqrt{\omega^2 - k^2}$.) 
The longitudinal susceptibility $\chi (\omega)$ is connected to the neutron scattering cross section
for experimental realizations in  quantum antiferromagnets. 

The second observable, the scalar susceptibility, is associated with two-point correlations
of the O($N$) invariant amplitude-squared of the order parameter $\sum_{\alpha=1}^N \phi_\alpha^2$ (in subsequent expression we will
omit the summation sign, and use an implicit Einstein convention for summation over repeated indices).
Specifically, we consider the response function
\beq
S (p) = \frac{1}{N}\int d^3 x \, e^{-i \bp \cdot \bx} \left[ \langle \phi_\alpha^2 (\bx) \phi_\beta^2 (0) \rangle -  \langle \phi_\alpha^2 (\bx) \rangle \langle \phi_\beta^2 (0) \rangle\right]
\eeq
and its analytic continuation $S(\omega)$ after $p \rightarrow - i \omega$.  
Our main results concern the structure of the function $S(\omega)$ in the vicinity of the
quantum critical point.  The recent experiments on ultracold bosons in optical lattices \cite{endres} measure the imaginary part, $S''(\omega)$, by determining the energy absorption rate under periodic oscillations in the strength of the lattice potential, which induce an oscillation in the hopping matrix element of the bosons.

The susceptibility $\chi (\omega)$ was computed in Refs.~\onlinecite{csy,ssrelax,zwerger}. 
In the Goldstone phase, 
it was found that there was a zero frequency peak in the imaginary part $\chi'' (\omega)$, with $\chi'' (\omega) \sim 1/\omega$,
and no finite $\omega$ peak which could indicate the energy of the Higgs excitation. When considered as an analytic function of
a complex frequency $\omega$, the retarded response function $\chi (\omega)$ was found to have a pole in the lower-half plane
at a purely imaginary frequency, indicating that the Higgs mode was purely relaxational. We will compute the leading $1/N$ corrections
to these results in our paper. We continue to find that $\chi'' (\omega ) \sim 1/\omega$ at small $\omega$. 
However, a qualitatively new feature
does emerge in the location of the pole
in the lower-half complex plane: it now appears at a frequency which has a non-zero real part, indicating an oscillatory component to the Higgs excitation.

In the Goldstone phase, the response function $S(\omega)$ also has a pole in the lower-half frequency plane at precisely the same location as that for
$\chi (\omega)$. However, now the oscillatory nature of the pole has a more visible signature on the real frequency axis.
Previous work \cite{csy,paa} found that the imaginary part behaves as $S'' (\omega) \sim \omega^3$ at small $\omega$.
At larger $\omega$, no peak-like structure was found in the $N=\infty$ computation in the strict scaling limit; this scaling limit
is taken by including only the single relevant perturbation away from the CFT3. Reference \onlinecite{paa} carried out computations
beyond the scaling limit at $N=\infty$, and found a finite $\omega$ peak, representing the first signature of a Higgs-like excitation.
In the present paper, we will compute $S(\omega)$ to order $1/N$. We find evidence for a finite-$\omega$ peak
in $S'' (\omega)$ even in the strict scaling limit. 
The issue of whether the peak in $S''(\omega)$ appears in the scaling limit, or not, is important because it determines the manner in which
the peak frequency, $\omega_p$ vanishes upon approaching the quantum critical point. A peak obtained in the scaling limit 
will have $\omega_p \sim (g_c - g)^\nu$, where $\nu$ the usual correlation length exponent, and $g$ is the coupling constant
tuning across the quantum critical point at $g=g_c$ (here the dynamic critical exponent $z=1$). 
Different behavior is expected for a peak which appears only because of corrections to scaling: in this case we 
can estimate from the results in Ref.~\onlinecite{paa} that $\omega_p \sim (g_c - g)^{\nu (1-\varpi)}$, 
where $\varpi$ is the leading correction to scaling exponent at the Wilson-Fisher fixed point. \cite{brezin}  Another key difference between the two types of peak is found in the behavior of the width of the peak, $\gamma$, as one approaches the critical point.  For a peak obtained in the scaling limit, the ratio $\gamma/\omega_p$ is fixed, and hence $\gamma$ must necessarily go to zero at the critical point. In contrast, for a peak which appears only due to corrections to scaling, $\gamma$ will in general not vanish at the critical point.

While our paper was being completed, we learnt of the numerical study of Pollet and Prokof'ev.\cite{pollet}
They found a peak in $S''(\omega)$ that appeared to vanish as $\omega_p \sim (g_c - g)^\nu$; 
this is further evidence that a peak is present already in the scaling limit.

We will present our scaling limit results for $S(\omega)$ (and $\chi (\omega)$) 
in terms of universal scaling functions. As in the analysis of Ref.~\onlinecite{solvay} for the 
quantum phase transition in the 
$d=3$ coupled-dimer antiferromagnet, it is useful to normalize the universal structure in the energy scale by comparing
the response functions on systems located symmetrically on opposite sides of the critical point. Thus let us take a point
in the gapped phase with O($N$) symmetry preserved with $g=g_+ > g_c$. Then its partner point with
$g=g_- < g_c$ is determined by
\beq
g_+ - g_c = g_c - g_- \label{flip}
\eeq
Let $\Delta$ be the {\em exact\/} energy gap to single-particle excitations in the symmetric phase at $g=g_+$;
consequently there is a pole in $\chi (\omega)$ at the real frequency $\omega=\Delta$ for $g=g_+$:
\beq
\frac{\omega_{\rm pole}}{\Delta} = 1 \quad , \quad g=g_+ , \label{realpole}
\eeq
and this defines the value of $\Delta$.  
By the definition of the critical exponent $\nu$,  $\Delta \sim (g_+- g_c)^\nu$ upon approaching $g_c$. 
Note that our definition of $\Delta$ is such that the threshold of $S''(\omega)$ in the symmetric phase occurs at $2\Delta$; for applications to the superfluid-insulator transition, $\Delta$ is the gap in the single boson Green's function in the Mott phase. We will use $\Delta$
to set the energy scale of {\em all\/} excitations, including those at $g=g_-$. 

For $g=g_-$, a key signature of the retarded response functions $\chi (\omega)$ and $S(\omega)$ is their common lower-half complex plane pole.
Its location, $\omega_{\rm pole}$, is completely determined by $\Delta$, and we find
\beq
\frac{\omega_{\rm pole}}{\Delta} = - i \, \frac{4}{\pi} + \frac{1}{N} \left(\frac{16\left(4+\sqrt{2}\ln \left(3-2\sqrt{2}\right)\right)}{\pi^2} + 2.46531203396 \, i \right) + \mathcal{O} (1/N^2) \quad , \quad g=g_- . \label{lhppole}
\eeq
The purely imaginary $N=\infty$ result was obtained earlier\cite{csy}. Note the key appearance of a real part $\sim 2.443216943237169/N$ in the
$1/N$ correction, which is a measure of the oscillatory component of the Higgs excitation.

We have claimed above that $\chi (\omega)$ and $S(\omega)$ have the same pole structure in the lower-half complex frequency plane
for $g=g_-$.
This is a consequence of the broken O($N$) symmetry, which mixes together the fluctuations of the longitudinal component and the amplitude-squared
of the order parameter. This will be evident from the structure of our analysis in Section~\ref{sec:gold}. 
In contrast, in the symmetric gapped phase $g=g_+$, these two observables have very different analytic structures. 
While $\chi (\omega)$ has a pole
on the real frequency axis, as in Eq.~(\ref{realpole}), the singularity of $S(\omega)$ is quite distinct.
This response function is in the two-particle
sector, and so for $g=g_+$ it has a threshold singularity at $\omega = 2\Delta$, as we noted above. 
We will show that this threshold singularity is of the form
\beq
S'' (\omega ) =  \frac{\widetilde{\mathcal{A}}\,\Delta^{3-2/\nu}}{\ln^2 \displaystyle \left(\frac{4\Delta}{(|\omega|- 2 \Delta)} \right)} \,
\mbox{sgn} (\omega) \, \theta (|\omega| - 2 \Delta) \quad , \quad g=g_+. \label{threshp}
\eeq
Here $\widetilde{\mathcal{A}}$ is a non-universal cut-off dependent prefactor that depends upon the precise definition of $\phi_\alpha$, and that
can be replaced by a constant in the vicinity of the critical point.

We also compute the universal structure of the amplitude-squared response function, $S(\omega)$, on the real frequency axis
in the Goldstone phase with $g=g_-$. 
Here, there is no threshold, and we have $S'' (\omega) \sim \omega^3$ as noted above.
More precisely, we find
\beq
S'' (\omega ) = \mathcal{A} \, \Delta^{-2/\nu} \omega^3 \quad \mbox{for} \quad \omega \rightarrow 0, ~g=g_-, \label{omega3}
\eeq
where $\mathcal{A}$ is another non-universal cut-off dependent constant prefactor, and $\Delta$ is defined as always by Eq.~(\ref{realpole}).
The recent observation of the low frequency spectrum in ultracold atom experiments \cite{endres} 
are consistent with the scaling in Eq.~(\ref{omega3}). The ratio of the amplitudes of the low frequency singularities
at $g_\pm$ in Eqs.~(\ref{threshp}) and (\ref{omega3}) are universal, and we find to order $1/N$ [from Eqs.~(\ref{tAval}) and (\ref{Aval})] that
\beq
\frac{\widetilde{\mathcal{A}}}{\mathcal{A}} = 32 \left( 1 + \frac{4.099}{N} \right) + \mathcal{O} (1/N^2).
\eeq

For general $\omega$, the scaling limit structure of $S(\omega)$ is (see the end of Sec.~\ref{sec:gen}) 
\beq
S (\omega) \sim \mbox{constant} + \Delta^{3-2/\nu} \Phi (\omega/\Delta), \label{Phiscale}
\eeq
where $\Phi$ is a universal function which we will compute here to order $1/N$;
the missing proportionality constant in Eq.~(\ref{Phiscale}) can be fixed by demanding consistency with Eqs.~(\ref{threshp}) and (\ref{omega3})
for $g=g_\pm$, and the additive constant is a non-universal real number.
We will find evidence that 
in the Goldstone phase ($g=g_-$),
the function $\Phi$ has a peak
as a function of $\omega$, which can be taken as an additional signature of the Higgs mode in two spatial dimensions.
For small $\omega$, the corrections to Eq.~(\ref{omega3}) are in Eq.~(\ref{Phires}) and these are schematically 
\beq
S'' (\omega)  \sim \omega^3 + \omega^5 \log (\omega). 
\eeq
This non-analytic correction is significant, and a signature of the strong coupling to the gapless modes. Associated non-analytic terms 
are responsible for the real part of the pole of $S(\omega)$ in Eq.~(\ref{lhppole}), as will become evident from the discussion 
in Section~\ref{sec:conc}.

We will begin by setting up the general formalism for the $1/N$ expansion for the Wilson-Fisher fixed point in Section~\ref{sec:gen}.
We will show here that the correlators of $\phi_\alpha^2$ can be efficiently computed by relating them to the correlators
of an auxillary scalar field. We will then apply this method to computations in the gapped phase, $g=g_+$, in Section~\ref{sec:symm}.
Our main results for the amplitude excitations of the Goldstone phase, $g=g_-$, appear in Section~\ref{sec:gold}. We present a unified review of our
results in Section~\ref{sec:conc}.

\section{General formalism}
\label{sec:gen}

In this section, we set up some general formalism for the O($N$) model. We will be interested in the field theory
\beq
Z_u [J] = \int \mathcal{D} \phi_\alpha \exp \left( -
\int_\bx  \left[ \frac{1}{2} (\partial \phi_\alpha)^2  + \frac{u}{2N} \left( \phi_\alpha^2  - N/g\right)^2 
+ J \phi_\alpha^2 \right] \right) \label{Zu}
\eeq
where $\int_\bx \equiv \int d^3 x$, $u$ is a quartic non-linearity, and $g$ is the tuning parameter across the transition.
We have included a  spacetime dependent source for the amplitude-squared mode, $J (\bx) $, and will eventually set $J(\bx)=0$ after taking 
functional derivatives of $Z_u (J)$. Note that the partition function in Eq.~(\ref{Zu}) is just the standard $\phi^4$ field theory of the
Wilson-Fisher fixed point, in an unconventional notation; the notation is designed to simplify the large $N$ limit,
and to make a natural connection to the fixed length non-linear sigma model in the limit $u \rightarrow \infty$.

We decouple the quartic term, and write the partition function as
\begin{equation}
Z_u [J] = \int \mathcal{D} \phi_\alpha \mathcal{D} \widetilde{\lambda} \exp \left( -
\frac{1}{2} \int_\bx  \left[ (\partial \phi_\alpha)^2
 + \frac{i}{\sqrt{N}} \widetilde{\lambda} \left(\phi_\alpha^2 - N/g\right) + \frac{\widetilde{\lambda}^2}{4 u} + 2 J \phi_\alpha^2 \right] \right) \label{Zu1}
\end{equation}
In the large $N$ limit, we perform the Gaussian functional integral over $\phi_\alpha$ (at $J=0$), and expand the fluctuations
in $\widetilde{\lambda}$ about the saddle point of the effective action. In the symmetric phase, the saddle point is at
$i \widetilde{\lambda}/\sqrt{N} = r $, where
\beq
\frac{1}{g} + \frac{r}{2u} = \int_\bp \frac{1}{p^2 + r}, \label{gval}
\eeq
where $\int_\bp \equiv  \int d^3 p /(8 \pi^3)$.
The nature of the saddle point in the Goldstone phase will be discussed later.
So we write $\widetilde{\lambda} = - i \sqrt{N} r + i 2 \sqrt{N} J + \lambda$, and completely integrate out the $\phi_\alpha$ field.
Then, the effective action for $\lambda$ is
\bea
Z_u [J]  &=& \int \mathcal{D} \lambda \exp \left(-\mathcal{S}_J - \mathcal{S}_0 - \mathcal{S}_1 \right)\nn
\mathcal{S}_0 &=& 
\frac{1}{2} \int_\bp  \left( \frac{\Pi(p,r)}{2} + \frac{1}{4u} \right) \lambda^2  \nn
\mathcal{S}_J &=& - \frac{Nr}{2g} - \frac{N r^2}{8u} + \int_x \left[ J \  \frac{\left(N r + 2 N u/g + i \sqrt{N} \lambda\right)}{2 u} - \frac{N J^2}{2 u} \right], \label{zuj}
\eea
where $\mathcal{S}_1$ involves cubic and higher order terms in $\lambda$ alone, and is specified in Appendix~\ref{app:lambda};
we note that $\mathcal{S}_1$ is independent of $u$.
We show in Appendix~\ref{app:crit} that $r$ plays the role of the tuning parameter away from the quantum critical point, with
$r \sim (g - g_c)^2$ for $g>g_c$.

Now we can take functional derivatives with respect to $J$, and obtain the two-point correlator of $\phi_\alpha^2$
at $J=0$ as
\beq
\langle \phi_\alpha^2 (\bx) \phi_\beta^2 (0) \rangle - \langle \phi_\alpha^2 \rangle^2 = \frac{N}{u} \delta^d (\bx) - \frac{N}{4u^2} \left[
\langle \lambda (\bx) \lambda (0) \rangle - \langle \lambda \rangle^2 \right] \label{amplam}
\eeq
So we see that the amplitude-squared correlator, $S(p)$,  is simply proportional to 
the two-point correlator of the $\lambda$ field. Specifically, we will identify
\beq
S(\omega) \equiv \mbox{constant} - G_{\lambda\lambda} (\omega), \label{SG}
\eeq
where the right hand side contains the connected $\lambda$ Green's function, and we are not interested in the additive
non-universal constant.
The function $G_{\lambda\lambda} (\omega)$ will be computed in this paper in the $1/N$ expansion;
this expansion can be carried out at any value of $u$, including the theory 
in which we impose the fixed length limit by taking the $u \rightarrow \infty$ limit. 
We also note that the identity (\ref{amplam}) also holds in the Goldstone phase.
Having proved Eq. (\ref{amplam}), we will set $J=0$ is all subsequent computations.

In the symmetric phase, $g>g_c$, we can write the $\lambda$ Green's function as
\beq
[G_{\lambda\lambda} (p)]^{-1} = [G_{\lambda\lambda}^{0} (p)]^{-1} - \Sigma_{\lambda\lambda} (p).
\eeq 
The bare propagator is 
\beq
G_{\lambda\lambda}^{0} (p) = \frac{2}{\Pi (p,r) + 1/(2u)}, \label{l0}
\eeq
where
\begin{eqnarray}
\Pi (p,r) = \int_\bq \frac{1}{(q^2+r)((\bp + \bq)^2 + r)} &=& \frac{1}{4 \pi p} \tan^{-1} \left( \frac{p}{2 \sqrt{r}} \right) \nn
&=& \frac{1}{8 \pi p i } \ln \left( \frac{2 \sqrt{r} + i p}{2 \sqrt{r} - i p} \right). \label{pires}
\end{eqnarray}
Note that $\Pi (p,r)$ diverges as $p, r \rightarrow 0$, and so the $1/(2u)$ term in Eq.~(\ref{l0}) can be neglected in this limit.
Because $u$ does not appear in the non-linear terms in $\mathcal{S}_1$, 
 we need to send $u \rightarrow \infty$ to obtain the scaling limit, and we will do so in most of our expressions;
 thus the scaling limit response function in the vicinity of the quantum critical point is the same for the ``soft spin'' (finite $u$)
 and fixed length ($u = \infty$) theories.
In terms of the expressions obtained in Appendix~\ref{app:lambda}, the expression for the self-energy to order $1/N$ is
\bea
\Sigma_{\lambda\lambda} (p) &=& -\frac{1}{2N} \int_\bk \left[K_3 (p,k,|\bp+\bk|) \right]^2 G_{\lambda\lambda}^0 (|\bk+\bp|) G_{\lambda \lambda}^0 (k)  \nn
&-& \frac{1}{2N} K_{3} (p,-p,0) G_{\lambda\lambda}^0 (0) \int_\bk K_3 (\bk,-\bk,0) G_{\lambda\lambda}^0 (k) \nn
&+& \frac{1}{6N} \int_\bk K_4 (\bp,\bk,-\bp,-\bk) G_{\lambda\lambda}^0 (k)
+ \frac{1}{3N} \int_\bk K_4 (\bp,-\bp,\bk,-\bk) G_{\lambda\lambda}^0 (k). \label{l1}
\eea

We can also obtain the expression for the $\phi_\alpha$ Green's function, $G(p)$, by similar methods. To order $1/N$ 
we obtain
\begin{eqnarray}
G^{-1} (p) &=& p^2 + r + \frac{1}{N} \int_\bq G_{\lambda\lambda}^0 (q)
\frac{1}{((\bp + \bq)^2 + r)} \nn
&~&~~~~- \frac{1}{2N} G_{\lambda\lambda}^0 (0) \int_{\bk, \bq}
\frac{1}{(k^2 + r)^2 ((\bk + \bq)^2 + r)} G_{\lambda\lambda}^0 (q). \label{gres}
\end{eqnarray}
We describe the structure of the critical singularities, as $g$ approaches
$g_c$, in the expressions in Eqs.~(\ref{l1}) and (\ref{gres})  in Appendix~\ref{app:crit}.

Before proceeding further, we give a brief derivation of Eq.~(\ref{Phiscale}), which plays a central role in our analysis.  Consider the $D=3$ dimensional statistical mechanics model defined by the partition function in Eq.~(\ref{Zu}), and define the free-energy density $f\equiv -{1\over V}\ln Z_u[J=0]$.  Then, taking two derivatives of $f$ with respect to $g^{-1}$ we obtain,
\bea
\frac{\partial^2 f}{(\partial g^{-1})^2 }=-uN+\frac{u^2}{V} \int_{\bx,\by}  \left[ \langle \phi_\alpha^2 (\bx) \phi_\beta^2 (\by) \rangle -  \langle \phi_\alpha^2 (\bx) \rangle \langle \phi_\beta^2 (\by) \rangle\right],
\eea
which is the $p=0$ Matsubara frequency correlation function $S(p=0)$, up to additive and multiplicative constants.  On the other hand, near the critical point $g_c$, $f$ can be split into regular and singular parts, $f=f_{reg}+f_{sing}$ where, according to the hyperscaling hypothesis, $f_{sing}\sim \xi^{-D}$.  Here, the correlation length $\xi\sim|g-g_c|^{-\nu}$ or, equivalently, $\xi\sim{\left|g_c^{-1}-g^{-1}\right|^{-\nu}}$.  By taking two derivatives of $f_{sing}$ with respect to $g^{-1}$, we conclude
\bea
S(p=0)\sim \xi^{-D+2/\nu} + ({\mathrm{regular\, constant}}).
\eea
Finally, since $\xi\propto 1/\Delta$ for a relativistic theory, and generalizing this result to arbitrary real frequencies, we obtain
\beq
S (\omega) \sim \Delta^{3-2/\nu} \Phi (\omega/\Delta)+  \mbox{constant},
\eeq 
which is Eq.~(\ref{Phiscale}).

\section{Spectral functions in the symmetric phase}
\label{sec:symm}

This section limits attention to the gapped phase where the O($N$) symmetry is preserved, present for $g>g_c$.
The dominant singularity in the $\phi_\alpha$ propagator is the pole corresponding to the $N$-fold degenerate
particle excitation. Let us determine the position and residue of this pole to order $1/N$.

The quasiparticle pole is determined by the solution of $G^{-1} (p)=0$ where from (\ref{chi})
\bea
G^{-1} (p) &=&  p^2 + r + \frac{1}{N \pi p} \int_0^\Lambda dq \frac{q^2}{\tan^{-1} (q/(2 \sqrt{r}))} \left[
\ln \left(\frac{ (p + q)^2 + r}{(p-q)^2 + r}\right) - \frac{4 p q}{q^2 + 4r} \right] 
\eea
So the pole is at $p^2 = - \Delta^2$ where
\beq
\Delta^2 = r  + \frac{1}{N \pi} \int_0^\Lambda dq \frac{q^2}{\tan^{-1} (q/(2 \sqrt{r}))} \left[
\frac{2}{\sqrt{r}} \tan^{-1} \left( \frac{2 \sqrt{r}}{q} \right)   - \frac{4 q}{q^2 + 4r} \right]
\eeq
The integral yields
\beq
\Delta^2 = r \left[ 1 + \frac{32 }{3 \pi^2 N} \left( \ln \left( \frac{\Lambda^2}{r} \right)  + 0.50260886161586 \right)\right]. \label{Delta1}
\eeq
This is the needed connection between the renormalized energy gap, $\Delta$, and the bare tuning parameter $r$.
Recall that we showed in Appendix~\ref{app:crit} that $r \sim (g-g_c)^2$, and so
 $\Delta \sim (g-g_c)^{\nu}$, where
\beq
\nu = 1 - \frac{32}{3 \pi^2 N}. \label{nu}
\eeq
Also, we have \cite{csy} for the wave-function renormalization $\mathcal{Z}$:
\bea
\frac{1}{\mathcal{Z}} &=& \left.  \frac{d G^{-1} (p)}{dp^2} \right|_{p^2=-\Delta^2} \nn
&=&  1 + \frac{1}{N \pi r} \int_0^\Lambda dq \frac{q^2}{\tan^{-1} (q/(2 \sqrt{r}))} 
\left[\frac{1}{\sqrt{r}}  \tan^{-1} \left( \frac{2 \sqrt{r}}{q}\right) - \frac{q}{q^2 + 4r}  - \frac{1}{q} \right] \nn
&=& 1 + \eta \left(\ln \left( \frac{\Lambda}{\sqrt{r}} \right) -1.78297175565687  \right) \label{Zval}
\eea
where the critical exponent $\eta$ is given by Eq.~(\ref{eta}).

\subsection{Threshold singularity}

As we discussed in Section~\ref{sec:intro}, the low frequency singularity in $G_{\lambda\lambda}$ is a threshold at a frequency $2 \Delta$.
So we consider the non-analytic terms in  $G_{\lambda\lambda} (p)$ at $p = - i (2 \Delta +  \omega) $ as $\omega \rightarrow 0$.

At $N = \infty$ we have
\bea
G_{\lambda\lambda}^0 (p) &=& \frac{2}{\Pi (- 2 i \sqrt{r} - i \omega, r)} \nn
&=& \frac{32 \pi \sqrt{r} }{\ln \left( \displaystyle \frac{4 \sqrt{r}}{-\omega} \right)}
\eea
So near threshold we have
\beq
\mbox{Im} G^0_{\lambda\lambda} (p) = \frac{32 \pi^2 \sqrt{r}}{\ln^2 (4 \sqrt{r}/|\omega|)}
\theta (\omega)
\eeq
The $1/\ln^2$ prefactor is easy to understand. We are considering the creation of two slowly moving particles in two dimensions with a
$k^2$ dispersion at low momenta; such particles have a well-known logarithmic singularity in their $T$-matrix for short-range interactions.\cite{ssbook}
This argument also implies that the $1/\ln^2$ prefactor will be robust at higher orders at $1/N$, as we indeed find below.

At order $1/N$, we compute the threshold singularity in $G_{\lambda\lambda}$ from Eq.~(\ref{l1}). Notice that the second and fourth terms in  Eq.~(\ref{l1}) can be
absorbed into a self-energy renormalization of the $\phi_\alpha$ propagator, which only contributes via the quasiparticle
wavefunction renormalization to the threshold singularity. So for the purposes of the threshold singularity we can write
\beq
G_{\lambda\lambda} (p) = \frac{2}{\mathcal{Z}^2 \Pi(p,\Delta^2)}
- \frac{8}{N\Pi^2 (p,r)} \int_\bk \frac{\left[K_3 (p,k,|\bp+\bk|) \right]^2}{ \Pi (|\bk+\bp|,r) \Pi (k,r)} 
+ \frac{4}{3N \Pi^2 (p,r)} \int_\bk \frac{K_4 (\bp,\bk,-\bp,-\bk)}{\Pi (k,r)}. \label{Gth}
\eeq
It remains to evaluate the $\bk$ integrals in the limit $\omega \rightarrow 0$. We present details of this evaluation
in Appendix~\ref{app:thresh}. The final result can be written as
\beq
\mbox{Im} G_{\lambda\lambda} (p) = \frac{\widetilde{\mathcal{A}} \, \Delta^{3-2/\nu}}{\ln^2 (4\Delta/|\omega|)}
\theta ( \omega) \label{thresh}
\eeq
with the amplitude $\widetilde{\mathcal{A}}$ is that appearing in Eq.~(\ref{threshp}), and is given by
\beq
\widetilde{\mathcal{A}}  = 32 \pi^2 \Lambda^{2/\nu - 2} \left( 1 + \frac{1.481}{N} + \ldots \right). \label{tAval}
\eeq

\section{Spectral functions in the Goldstone phase}
\label{sec:gold}

We write $\phi_\alpha = (\sqrt{N} \sigma_0 + \sigma, \pi_i)$ with  $i = 1 \ldots N-1$,
where
\beq
\sigma_0^2 = \frac{1}{g} - \int_\bp \frac{1}{p^2}, \label{cond}
\eeq 
is the condensate at $N=\infty$. From Eqs.~(\ref{flip}) and (\ref{rgpi}), we can relate this to the value of $r$ at
the partner point in the symmetric phase
\beq
\sigma_0^2 = \frac{\sqrt{r}}{4\pi},
\eeq
where $r$ is related to the single particle gap in the paramagnet via Eq.~(\ref{Delta1}).  This yields a relation between $\sigma_0^2$ and the gap at the partner point in the symmetric phase:
\beq
\sigma_0^2 = \frac{\Delta}{4 \pi}  - \frac{32 \sigma_0^2}{3 \pi^2 N } \ln \left( \frac{\Lambda}{16 \sigma_0^2} \right)  - \frac{\mathcal{C} \sigma_0^2}{N},
\label{Delta}
\eeq
where $\mathcal{C} = 0.5326726500732776$.  This can also be written as
\beq
\sigma_0^2 = \frac{\Delta}{4 \pi}\left[1- \frac{\mathcal{C}}{N}   - \frac{32}{3 \pi^2 N } \ln \left( \frac{\pi \Lambda}{4\Delta} \right)  \right]
\label{Delta2}
\eeq
within the same order in $1/N$.

We can set
$\widetilde{\lambda} = \lambda$ in Eq.~(\ref{Zu1}), and
then, integrating over the $\pi_i$ we obtain the partition function 
\bea
Z &=& \int \mathcal{D} \sigma \mathcal{D} \lambda \exp \left( - \mathcal{S}_0 - \mathcal{S}_1 - \mathcal{S}_2 \right),\nn
\mathcal{S}_0 &=&
\frac{1}{2} \int_\bp  \left[ p^2 \sigma^2 + 2 i \sigma_0 \sigma \lambda + \frac{1}{2} \Pi(p,0) \lambda^2 \right],\nn
\mathcal{S}_2 &=& -\frac{i}{2\sqrt{N}} \left( \frac{1}{g} - \sigma_0^2 \right) \int_x \lambda - \frac{1}{4N} \int_\bp \Pi (p,0) \lambda^2 
+ \frac{i}{2\sqrt{N}} \int_x
  \lambda\, \sigma^2, \label{zhiggs}
\eea
where $\mathcal{S}_1$ is just as in Appendix~\ref{app:lambda} but with $K_{3,4}$ evaluated at $r=0$.
Now the connected $\sigma$-$\sigma$ Green's function is equivalent to the longitudinal susceptibility $\chi (\omega)$ introduced
in Sec.~\ref{sec:intro}:
\beq
\chi (\omega) \equiv G_{\sigma\sigma} (\omega). \label{chiGss}
\eeq

The bare connected Green's functions of Eq.~(\ref{zhiggs}) are
\bea
G_{\sigma\sigma}^0 (p) &=& \frac{\Pi (p,0)}{p^2 \Pi (p,0) + 2 \sigma_0^2},\nn 
G_{\lambda\lambda}^0 (p) &=& \frac{2p^2}{p^2 \Pi (p,0) + 2 \sigma_0^2}, \nn 
G_{\sigma\lambda}^0 (p) &=& \frac{-2 i\sigma_0}{p^2 \Pi (p,0) + 2 \sigma_0^2}.  \label{bareG}
\eea
The corresponding renormalized connected Green's functions require computations of the self-energies at order $1/N$, and these
are presented in Appendix~\ref{app:gold}. 

Using the value of $\Pi (p,0)$ in Eq.~(\ref{pires}), we see that all the Green's functions have poles at $p=-16 \sigma_0^2$,
which corresponds to the lower-half plane value $\omega = -i 16 \sigma_0^2$. After including the self-energies, this pole corresponds
to the point where the effective quadratic action for $\lambda $ and $\sigma$ has a zero eigenvalue. Equivalently, this is the condition 
that the determinant of the quadratic form vanish, i.e.,
\beq
\mbox{det} \left( \begin{array}{ccc}
p^2 - \Sigma_{\sigma\sigma} (p) &~& i \sigma_0 - \Sigma_{\sigma\lambda} (p) \\
i \sigma_0 - \Sigma_{\sigma\lambda} (p) &~& 1/(16 p) - \Sigma_{\lambda\lambda} (p) \label{det}
\end{array} \right)=0,
\eeq
where $\Sigma$ are self-energies computed in Appendix~\ref{app:gold}. Note that upon 
ignoring the self-energies, the solution of Eq.~(\ref{det}) is indeed $p = -16 \sigma_0^2$.
Rather than solving this equation with self-energies to next order in $1/N$, it is more convenient to locate
the pole by examining the zero of $1/G_{\sigma\sigma}$, which we will do in Sec.~\ref{sec:long}.

\subsection{Computation of $\Phi$}
\label{sec:Phi}

We can now use the results for the self energies computed 
in Appendix~\ref{app:gold} to obtain expressions for the connected $\lambda$ Green's functions to order $1/N$:
\bea
G_{\lambda\lambda} (p) &=& G_{\lambda\lambda}^{0} (p) +  \Sigma_{\lambda\lambda} (p) \left[G_{\lambda\lambda}^{0} (p) \right]^2
+ \Sigma_{\sigma\sigma} (p) \left[G_{\lambda\sigma}^{0} (p) \right]^2 + 2 \Sigma_{\lambda\sigma} (p) G_{\lambda\lambda}^{0} (p)G_{\lambda\sigma}^{0} (p) \nn
&=& \frac{16p^2}{p + 16 \sigma_0^2 }  + \frac{16 p^3}{N (p+ 16 \sigma_0^2)^2} 
+ \frac{1024 p^2 ( p + 8 \sigma_0^2)}{3 \pi^2 N (p + 16 \sigma_0^2)^2}  \ln \left( \frac{\Lambda}{16 \sigma_0^2} \right) 
\nn
 &~&~~~+ \frac{256 \sigma_0^2}{(p+ 16 \sigma_0^2)^2 N} \left[ p^2 F_{\lambda\lambda} (p)  - \sigma_0^4 F_{\sigma\sigma} (p) +2p \sigma_0^2 F_{\sigma\lambda} (p) \right] \nn
&~&~- \frac{256 \Lambda}{\pi^2 N}
+  \frac{8192 \sigma_0^2}{\pi^2 N} \ln \left( \frac{\Lambda}{16 \sigma_0^2} \right), \label{Glambda}
\eea
where the $F$ functions are defined in Eqs.~(\ref{Fdefs1})-(\ref{Fdefs3}).
The terms in the last line are $p$-independent real constants, and will be dropped since they do not contribute to $S''(\omega)$
upon using Eq.~(\ref{SG}). After this subtraction, it is notable that all the terms linear in $\Lambda$ have canceled out of this expression,
indicating that the $S(\omega)$ is universal near the quantum critical point.

In order to extract the universal scaling function $\Phi$ from $G_{\lambda\lambda}$, as introduced in 
Eq.~(\ref{Phiscale}), we first collect the $p$-independent terms appearing in the last line of Eq.~(\ref{Glambda}).  
These terms are explicitly real and can be absorbed into the constant in Eq.~(\ref{Phiscale}).   
We then use Eq. (\ref{Delta2}) to substitute $\sigma_0$ by $\Delta$, and find:
\bea
G_{\lambda\lambda} (p)&=&({\rm const})+\frac{16 p^2}{p+4 \Delta/\pi }
                         +\frac{16 p^2(p+4\mathcal{C}\Delta/\pi)}{N \left(p+4 \Delta/\pi\right)^2}
						+\frac{1024 p^2}{3 \pi ^2 N \left(p+4 \Delta/\pi\right)} \ln
						   \left(\frac{\pi  \Lambda }{4 \Delta }\right) \nn
                      &~&~~~   +\frac{4\Delta}{N \pi^3(p+4 \Delta/\pi)^2}  \left[16\pi^2 p^2 F_{\lambda \lambda}(p)-\Delta^2 F_{\sigma \sigma}(p)+8 \pi p \Delta F_{\sigma\lambda} (p) \right] 
\eea
Using
\bea
(\Delta/\Lambda)^{2-2/\nu}=1+\frac{64}{3\pi^2 N}\ln (\Delta/\Lambda),
\eea
this can be written as
\bea
G_{\lambda\lambda} (p)&=& ({\rm const})+\Delta \left(\Delta/\Lambda\right)^{2-2/\nu} H_\lambda(p/\Delta), \label{GtoH}
\eea
where
\bea
H_\lambda(p/\Delta)&=&\frac{16 p^2}{\Delta(p+4 \Delta/\pi) }
  						+\frac{16 p^2(p+4\mathcal{C}\Delta/\pi)}{N \Delta \left(p+4 \Delta/\pi\right)^2}
	 					+\frac{1024 p^2}{3 \pi ^2 N \left(p+4 \Delta/\pi\right)} \ln
						   \left(\frac{\pi}{4 }\right) \nn
						&~&~~ +\frac{4}{N \pi^3(p+4\Delta /\pi)^2}  \left[16\pi^2 p^2 F_{\lambda \lambda}(p)-\Delta^2 F_{\sigma \sigma}(p)+8 \pi p \Delta F_{\sigma \lambda}(p)\right]. \label{PhiLambdaEuclidean}
\eea
Importantly, notice that all the logarithmic dependence upon the cutoff $\Lambda$ has dropped out, 
yielding a universal expression for $H_\lambda$; this is a significant check on our computation.  
By comparing Eq.~(\ref{GtoH}) to Eqs.~(\ref{Phiscale}) and (\ref{SG}), we see that the scaling function $\Phi$ is
\bea
\Phi\left(\frac{\omega}{\Delta}\right)=-H_\lambda\left(\frac{-i\omega}{\Delta}\right). \label{HtoPhi}
\eea
Hence,  $\Phi$ is obtained from the analytic continuation of $H_\lambda$ to $p=-i\omega$. 

At small values of $z=\omega/\Delta,$ the expansion (\ref{FsmallP}) yields
\bea
\Phi\left(z\right) &=& 
\left(4\pi-\frac{25.898}{N}\right)z^2+i \left(\pi^2 - \frac{25.842}{N} \right) z^3+\left(-\frac{\pi^3}{4}+\frac{40.690}{N}\right)z^4 \nn 
&+& \left(-\frac{i \pi^4}{16}+\frac{50.584 i}{N}  - \frac{2i\pi^2}{3N} \ln (-i z) \right) z^5 +\mathcal{O}\left(z^6\right). \label{Phires}
\eea
Note that the imaginary part of $\Phi$ has a leading $\omega^3$ behavior.  This relies on a precise cancellation of a term $\sim \omega$, and is consistent with previous results \cite{csy,paa}.  The coefficient of $\omega^3$ is a measure of the decay rate of the Higgs excitation into two gapless Goldstone modes.  
Comparing the definition in Eq.~(\ref{omega3}) with Eq.~(\ref{Phires}) we have
\beq
\mathcal{A} = \Lambda^{2/\nu - 2}  \left(\pi^2 - \frac{25.842}{N} \right). \label{Aval}
\eeq
Note that there is a branch cut in Eq.~(\ref{Phires}) at $\omega=0$, as 
indicated to lowest order in $\omega$ by the term $\omega^5 \log{\omega}$.  This is a non-analytic 
consequence of the presence of gapless excitations. This same non-analytic term is responsible for the real part of the pole in Eq.~(\ref{lhppole}),
as is shown in Appendix~\ref{app:gold} and discussed in Section~\ref{sec:conc}.

We next evaluate $\Phi$ numerically at general values of $\omega$.  For this, we must perform an analytic continuation of the integrals $F$ in Eqs.~(\ref{Fdefs1}-\ref{Fdefs3}) to real frequency $p=-i\omega$.  A Wick rotation of these integrals  can be performed by taking $p=e^{-i\alpha}\omega$ and $k=e^{-i\alpha} \nu$, and varying $\alpha$ continuously from $\alpha=0$ (the Euclidean time result) to $\alpha=\pi/2$ (real time).  On general grounds, we expect not to cross any singularities upon this Wick rotation---indeed, this can be verified explicitly by inspection of the integrands.   Thus we interpret   $F(p=-i\omega)$ in Eqs.~(\ref{Fdefs1})-(\ref{Fdefs3}) as integrals along straight line segments on the complex plane, running from $k=0$ to $k=-i\omega$, and from $k=-i\omega$ to $k=-i\infty$.

\begin{figure}[!t]
\begin{center}
\includegraphics[width=12cm,angle=0]{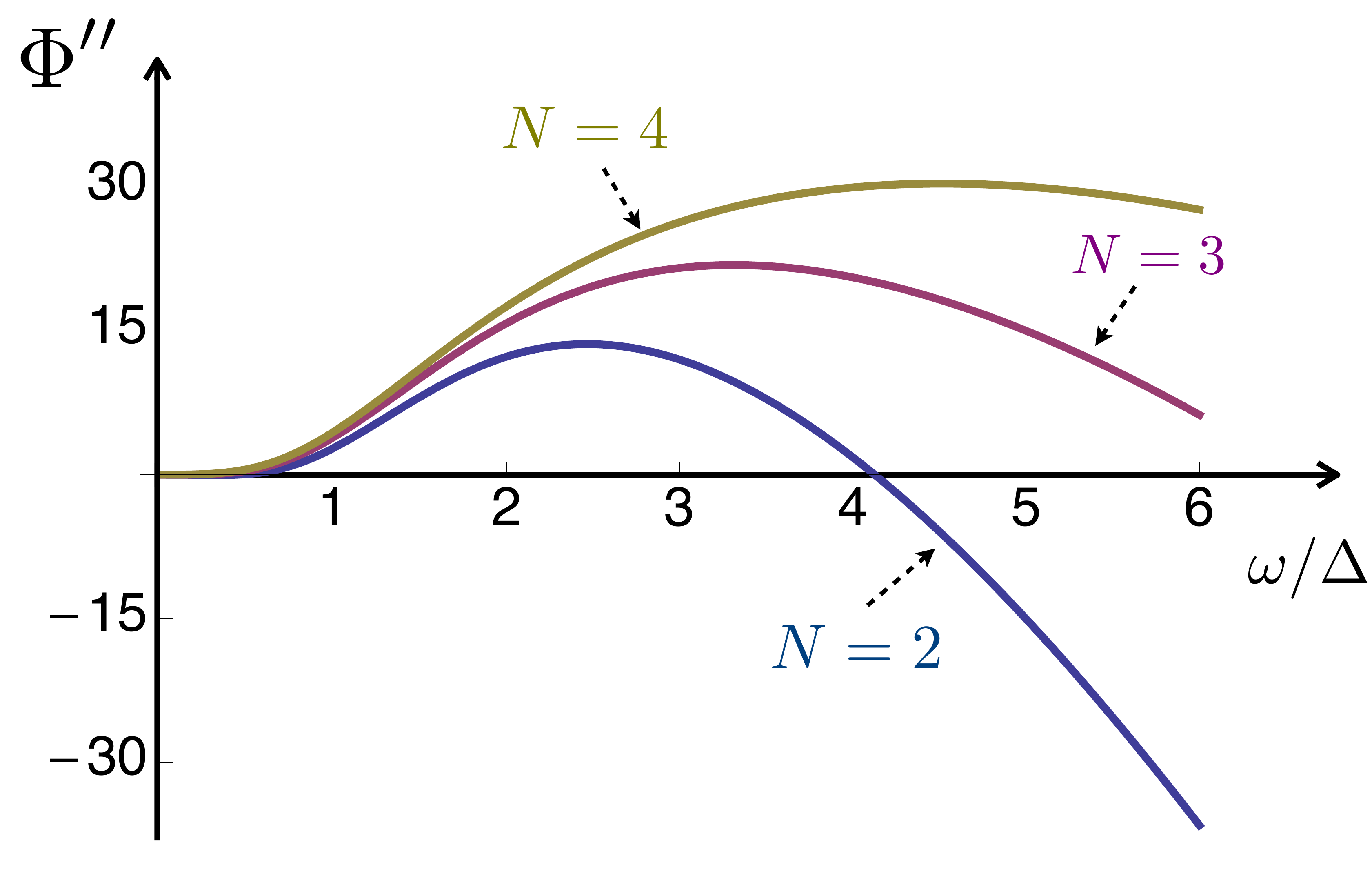}
\caption{Universal component of the scalar spectral function, $\Phi''(\omega/\Delta)$, for different values of $N$.  The change of sign in the $N=2$ curve at large $\omega/\Delta$ is an artifact of the approximation being used.}
\label{PhiLambda} 
\end{center}
\end{figure}

The universal scaling contribution to the scalar spectral function, $\Phi''(\omega/\Delta)$, is shown in Fig.~\ref{PhiLambda} for $N=2$, $N=3$, and $N=4$.   Note that $\Phi''$ has a peak in all cases.  This is one of the main results of our analysis.  It supports the appearance of the Higgs mode as a finite width peak in scalar measurements in the quantum critical region.  The presence of this peak is substantiated by recent experiments\cite{endres} and quantum Monte-Carlo simulations\cite{pollet} of the Mott-superfluid transition at integer filling of bosons, corresponding to the case $N=2$.

\begin{figure}[!t]
\begin{center}
\includegraphics[width=12cm,angle=0]{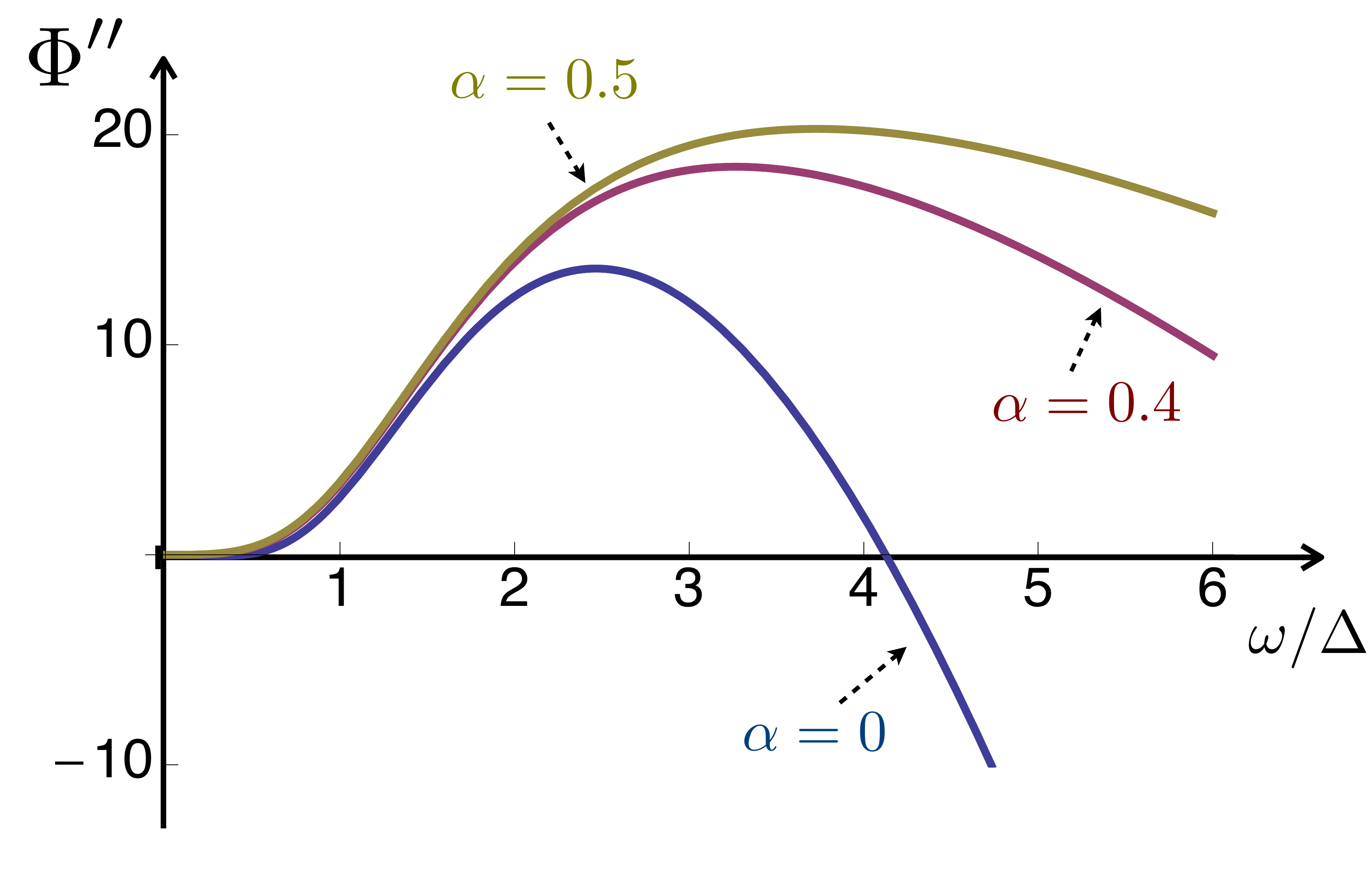}
\caption{Universal component of the scalar spectral function $\Phi(\omega/\Delta)$ for $N=2$, using for $N=2$, using the Pad\'e approximant Eq.~(\ref{PhiPade}) for different values of $\alpha$.  For $\alpha=0$, corresponding to Eq. (\ref{PhiTaylor}), $\Phi''$ becomes negative at very small and very large $\omega/\Delta$.  This unphysical result is absent at $\alpha=0.4$ and $\alpha=0.5$. }
\label{pade} 
\end{center}
\end{figure}

The $N=2$ curve in Fig.~\ref{PhiLambda} is negative at large $\omega$, and also at very low values of $\omega$, as can be inferred from the sign of $\mathcal{A}$ in Eq.~(\ref{Aval}) for $N=2$ (resulting in a very shallow negative region that cannot be discerned in the scale of Fig.~\ref{PhiLambda}).  These regions of negative $\Phi''$ are unphysical, since the spectral function must always be positive.  Note that the scaling function can be written as
\bea
\Phi(z)=\Phi_0(z)+\frac{1}{N}\Phi_1(z)+\mathcal{O}(1/N^2) \label{PhiTaylor}
\eea
The change of sign in $\Phi''$ occurs because the $1/N$ correction $\Phi_1$ can become dominant over the $N=\infty$ result, $\Phi_0$, both at very low and very large $z$.  This is an indication of the breakdown in the $1/N$ expansion at the current order in the case $N=2$. One possible method to avoid the unphysical regions is to introduce a Pad\'e approximant,
\bea
\Phi_{\rm{Pad\acute{e}}}(z)=\frac{\Phi_0(z)+\frac{1-\alpha}{N}\Phi_1(z)}{1-\frac{\alpha}{N}\Phi_1(z)/\Phi_0(z)}+\mathcal{O}(1/N^2) \label{PhiPade}
\eea
which is equivalent to Eq. (\ref{PhiTaylor}) to our working order in $1/N$.  Then, by choosing an appropriate value of $\alpha\ne 0$, one obtains a positive spectral function, as shown in Fig.~\ref{pade}.  For instance, in the small $z$ region, Eq.~(\ref{Aval}) would be replaced by,
\bea
\mathcal{A}_{\rm{Pad\acute{e}}}=\Lambda^{2/\nu - 2}  \frac{\pi^2 - 25.842(1-\alpha)/N}{1+ 25.842\alpha/(\pi^2 N)},
\eea
which matches Eq.~(\ref{Aval}) to our working order in $1/N$, yet is positive for all values $N\ge 2$ provided $\alpha>0.2366$.

The breakdown in the $1/N$ result in fact occurs for all $N$: at large enough $z$, $\Phi''$ becomes negative for any $N$, although the value of $z$ where this occurs increases with $N$. This means that there is significant ambiguity in the large-$z$ dependence of $\Phi''$, as seen in Fig.~\ref{pade}.  However, many features of $\Phi''$ are robust.  For example, $\Phi''$ displays a peak provided that the value of $\alpha$ is moderate.  For $\alpha=0$, the peaks in Fig.~\ref{PhiLambda} are located at $\omega_{\rm peak}\approx 2.5 \Delta$ for $N=2,$ $\omega_{\rm peak}\approx 3.5 \Delta$ for $N=3,$ and $\omega_{\rm peak}\approx 4.5 \Delta$ for $N=4.$   There is uncertainty in the locations of the peaks, since these do shift when $\alpha$ is varied.  However, the fact that the location of the peak in $\Phi''$ is monotonically increasing with $N$ is robust for any value of $\alpha$.

\subsection{Computation of longitudinal susceptibility}
\label{sec:long}

The longitudinal susceptibility $\chi (\omega)$ is related to $G_{\sigma\sigma}$ via Eq.~(\ref{chiGss}). 
From scaling arguments we expect this response function to satisfy
\bea
G_{\sigma\sigma}^{-1} (\omega) &=& \Delta^{2-\eta} \Phi_\sigma (\omega/\Delta). \label{Sigmascale}
\eea
We will next compute the universal function $\Phi_\sigma$ to order $1/N$. 

To order $1/N$,
\bea
\frac{1}{G_{\sigma\sigma} (p)} &=& \frac{1}{G_{\sigma\sigma}^{0} (p)} - \Sigma_{\sigma\sigma} (p) -  \Sigma_{\lambda\lambda} (p) 
\frac{\left[G_{\lambda\sigma}^{0} (p) \right]^2}{\left[G_{\sigma\sigma}^{0} (p) \right]^2}
 - 2 \Sigma_{\lambda\sigma} (p) \frac{G_{\lambda\sigma}^{0} (p)}{G_{\sigma\sigma}^{0} (p)} \nn
&=& p(p+ 16 \sigma_0^2) + \frac{16 p \sigma_0^2}{N} + \frac{8 p (p + 80 \sigma_0^2)}{3 \pi^2 N}  \ln \left( \frac{\Lambda}{16 \sigma_0^2} \right) \nn
 &~&~~~- \frac{\sigma_0^4}{N} \left[ F_{\sigma\sigma} (p)  -256 F_{\lambda\lambda} (p) + 32 F_{\sigma\lambda} (p) \right], \label{Gsigma}
\eea
where the functions $F$ are defined in Eqs.~(\ref{Fdefs1}-\ref{Fdefs3}). Using Eq. (\ref{Delta2}), we find:
\bea
\frac{1}{G_{\sigma\sigma} (p)}&=& p \left(p+4 \Delta /\pi\right)+ \frac{4(1-\mathcal{C}) p \Delta }{\pi N} +\frac{8 p (p+4 \Delta/\pi ) }{3 \pi^2 N}\log \left(\frac{\pi  \Lambda }{4 \Delta }\right)\nn
					&~&~~~ -\frac{\Delta^2}{16\pi^2  N}\left[ F_{\sigma\sigma}(p)-256 F_{\lambda \lambda}(p)+32 F_{\sigma \lambda}(p)\right] \nn
                      &=&\Delta^2(\Delta/\Lambda)^{-\eta} H_\sigma(p/\Delta), \label{SigmaToH}
\eea
where we have used Eq. (\ref{eta}) to write
\bea
(\Delta/\Lambda)^{-\eta}=1-\frac{8}{3\pi^2 N}\ln (\Delta/\Lambda)
\eea
and where we have introduced
\bea
H_\sigma(p/\Delta)&=&\frac{p \left(p+4 \Delta /\pi\right)}{\Delta^2}+ \frac{4(1-\mathcal{C}) p  }{\pi N \Delta} +\frac{8 p (p+4 \Delta/\pi ) }{3 \pi^2 \Delta^2 N}\log \left(\frac{\pi}{4}\right)   \nn
  &~&~~~ -\frac{1}{16\pi^2  N} \left[ F_{\sigma\sigma}(p)-256 F_{\lambda \lambda}(p)+32 F_{\sigma \lambda}(p)\right] \label{PhiSigmaEuclidean}
\eea
which is seen to be a universal function of $p/\Delta$.  Comparison with Eq.~(\ref{Sigmascale}) gives
\bea
\Phi_\sigma(\omega/\Delta)=H_\sigma(-i\omega/\Delta).
\eea
At small values of $z=\omega/\Delta$, we can use Eq.~(\ref{FsmallP}) to obtain
\bea
\Phi_\sigma(z)&=&-\left(\frac{4}{\pi}+\frac{0.5119}{N}\right)i z-\left(1-\frac{2.2811}{N}\right)
z^2\nn
&~&~~~+\left(\frac{0.2793 i}{N}-\frac{24i}{9\pi N}\ln (-i z)\right)
z^3+\mathcal{O}(z^4)
\eea
Note the linear dependence at low $z$, corresponding to the infrared divergence in the longitudinal susceptibility $\chi''(\omega)\sim 1/\omega$.\cite{csy,ssrelax,zwerger,paa}

\begin{figure}[!t]
\begin{center}
\includegraphics[width=12cm,angle=0]{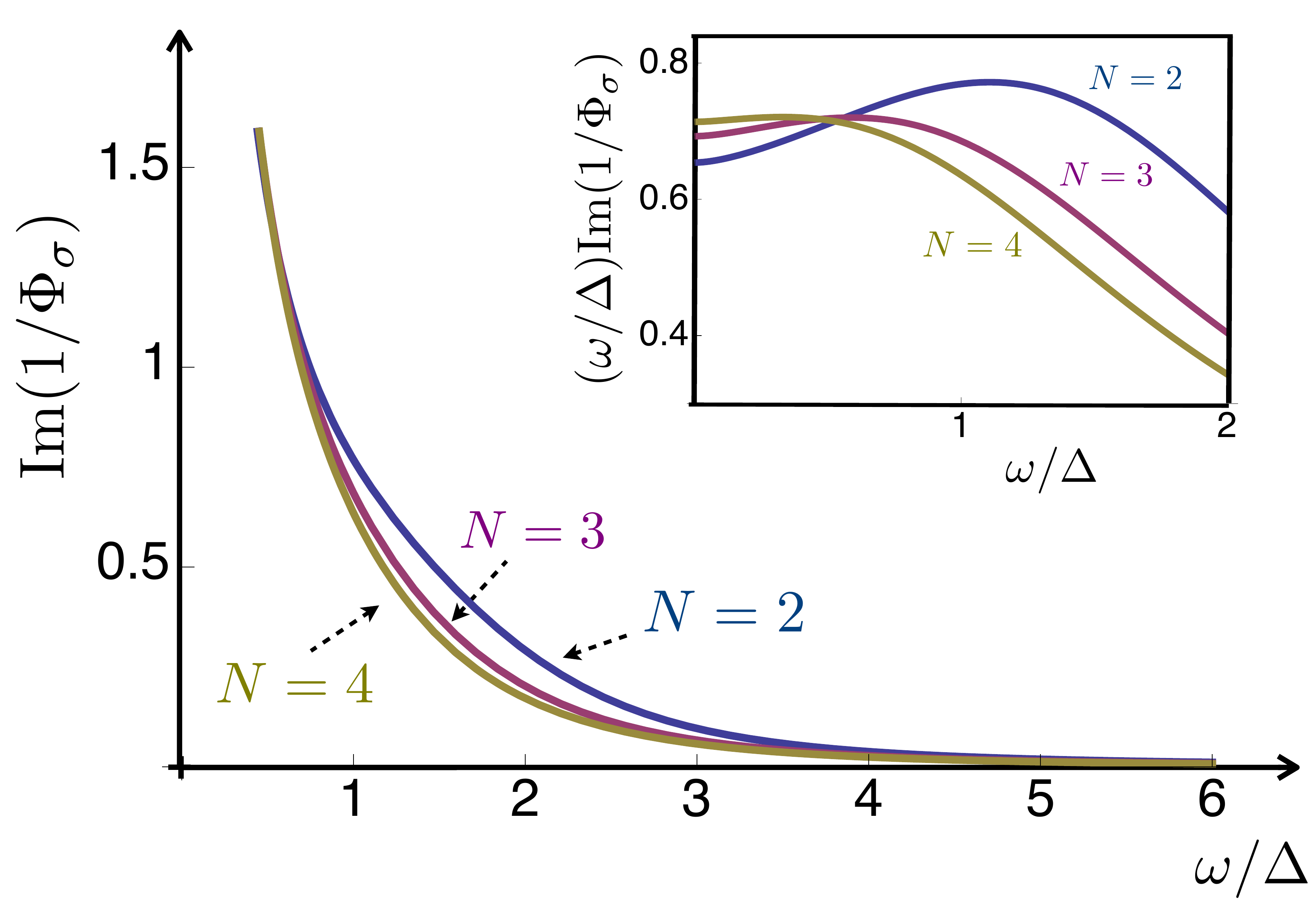}
\caption{The imaginary part of $1/\Phi_\sigma(\omega/\Delta)$, obtained from the analytic continuation of Eq.~(\ref{PhiSigmaEuclidean}) for $N=2$, $N=3,$ and $N=4$.  This corresponds to the universal scaling form of the longitudinal spectral function $\chi''(\omega)$. {\bf Inset:}  $\omega/\Delta \cdot {\rm Im}(1/\Phi_\sigma)$, for $N=2$, $N=3,$ and $N=4$.} 
\label{PhiSigma} 
\end{center}
\end{figure}

We compute $\Phi_\sigma$ numerically performing an analytic continuation as before.  Figure \ref{PhiSigma} shows the imaginary part of $1/\Phi_\sigma$ for the cases $N=2$, $N=3$, and $N=4.$  This corresponds to the universal scaling form of $\chi''(\omega)$.  In all cases, the curves are dominated by the infrared divergence $\sim 1/\omega$. In order to suppress this infrared divergence and highlight the features appearing at intermediate values of $\omega/\Delta$, we plot  $\omega/\Delta \cdot {\rm Im}(1/\Phi_\sigma)$ in the inset of Fig.~\ref{PhiSigma} .

Equations (\ref{SigmaToH}) and (\ref{PhiSigmaEuclidean}) display a key feature of the longitudinal Green's function, which first appears at our current working order, $\mathcal{O}(1/N)$.  For $N=\infty$, the longitudinal Green's function has poles at $p=0$ and $p=-4\Delta/\pi.$  The latter pole, which lies on the negative $p$ axis, corresponds to an overdamped response with purely imaginary frequency $\omega=-4i\Delta/\pi$.  Remarkably, at $\mathcal{O}(1/N)$, this pole acquires a real component.  The appearance of a real part in $\omega$ leads to a real-time longitudinal response which has both decaying and oscillatory components.
  
In order to find the shift in the location of this pole, we 
need to solve Eq.~(\ref{det}) to order $1/N$. Equivalently, we can
look for solutions to $H_{\sigma}(p=-4\Delta/\pi + \delta p)=0$, where $\delta p=\mathcal{O}(1/N)$.  This gives
\bea
\delta p =-\frac{4 \Delta}{ N \pi}\left\{1-\mathcal{C}+ \frac{1}{256}\left(F_{\sigma \sigma}+32  F_{\sigma \lambda}-256  F_{\lambda \lambda} \right) \right\}+\mathcal{O}(1/N^2)
\eea
where the functions $F$ are to be evaluated at $p=-4\Delta/\pi$.  This is carried out in Eq.~(\ref{poleshift}), where it is found that  $F_{\lambda\lambda}(-\Delta/\pi)$ has an imaginary component -- this is the source of the real frequency shift in the pole, $\omega_{\rm pole}=i (-4\Delta/\pi+\delta p)$.  We find
\beq
\frac{\omega_{\rm pole}}{\Delta} = - i \, \frac{4}{\pi} + \frac{1}{N} \left(\frac{16\left(4+\sqrt{2}\log \left(3-2\sqrt{2}\right)\right)}{\pi^2} + 2.46531203396 \, i \right) + \mathcal{O} (1/N^2)\label{poleposition}
\eeq
The real part is $16\left(4+\sqrt{2}\log \left(3-2\sqrt{2}\right)\right)/(N\pi^2) \sim 2.443216943237169/N$.  This pole, which is most easily extracted by studying the longitudinal susceptibility, will also appear in the scalar response function $\Phi(z)$, as discussed below.

\section{Conclusions}
\label{sec:conc}

Our paper has studied the nature of Higgs excitations of the relativistic O($N$) model in 2+1 dimension in the Goldstone
phase in the vicinity of the Wilson-Fisher fixed point.\cite{wf} It is remarkable that after nearly 40 years of study, 
zero temperature characteristics of this venerable fixed point
have still remained uncovered.

Our main results were for the spectral function of the amplitude-squared of the order parameter that is measured in
recent experiments on the superfluid-insulator quantum phase transition in Ref.~\onlinecite{endres}, and in numerical simulations
in Ref.~\onlinecite{pollet}. We expressed our results
in terms of
the universal complex scaling function $\Phi$ in Eq.~(\ref{Phiscale}), which measures frequency in terms of a single 
energy scale $\Delta$ which vanishes at the critical point with exponent $\nu$. We obtained information on distinct features
of this function: it was expanded in powers of $1/N$ in Section~\ref{sec:Phi}, expanded at low frequencies in Eq.~(\ref{Phires}),
and we determined the location of the pole in the lower-half complex frequency plane in Section~\ref{sec:long}.
The pole is specified in Eq.~\ref{lhppole}: note that both its real and imaginary parts are proportional to $\Delta$, and so the width
of the Higgs particle is of the same order as its energy.

\begin{figure}[!t]
\begin{center}
\includegraphics[width=5cm,angle=0]{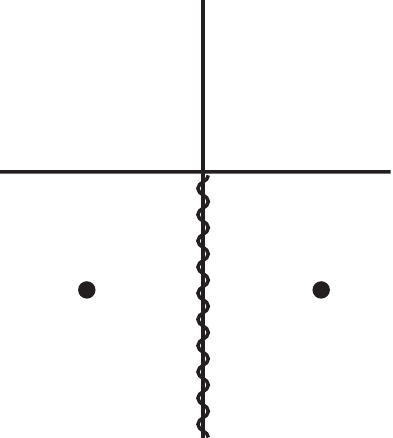}
\caption{Analytic structure of the {\em retarded\/} Green's function $\Phi (z)$ in Eq.~(\ref{phistruct}).
This is analytic and unique in the upper-half plane, but its structure in the lower-half plane depends upon the choice of a branch cut
originating at $z=0$.
 If we analytically continue from the upper-half plane, we find a pole
in the lower-half plane, but its location depends upon the manner in which we encircle $z=0$. The figure shows the choice of the branch cut
along the negative imaginary axis. Then the pole is at Eq.~(\ref{poleposition}) or at the partner point with the opposite sign for the real part.
Note the similarity to the analytic structure of the strongly coupled gapless systems studied using the AdS/CFT correspondence
in Refs.~\onlinecite{ads2,denef,hofman}.}
\label{fig:pole} 
\end{center}
\end{figure}

Here we combine these results to obtain useful interpolating forms for comparing with numerical or experimental data.
We consider $\Phi$ as function of the complex variable $z= \omega/\Delta = i p /\Delta$. An explicit form with all the necessary
analytic features is
\beq
\Phi (z) = \frac{z^2}{1 +  i c_1 z + c_2 z^2 +i c_3  z^3 + i c_4 z^3 \ln (-i z)}+c_0\label{phistruct}
\eeq
where $c_{0,1,2,3,4}$ are real constants. The constraints on these
constants are:
\begin{itemize}
\item
$\Phi (z)$ is real on the imaginary axis in the upper-half plane, $z=i p/\Delta$ with $p>0$.
\item $\Phi (z)$ is analytic in the upper-half plane.
\item There is a branch cut in $\Phi (z)$ starting from $z=0$, we take this cut to lie in the lower-half plane.
Upon analytically continuing $\Phi(z)$ clockwise from the imaginary axis in the upper-half plane
we find a single pole in the lower-half plane. The position of this pole should coincide with Eq.~(\ref{lhppole}).
On the other hand, if we analytically continue $\Phi(z)$ anticlockwise from the upper-half plane we again find only a single
pole in the lower-half plane, but its real part has changed sign. One convenient choice is to place the branch cut along the lower
imaginary axis, leading to the analytic structure shown in Fig.~\ref{fig:pole}. This analytic structure is remarkably similar to that
found in recent studies using the AdS/CFT correspondence.\cite{ads2,denef,hofman} These studies describe a particle interacting strongly with a gapless
continuum of excitations, and so bears a similarity to our study of the Higgs particle interacting strongly with the Goldstone excitations.
\item The small $z$ expansion of $\Phi (z)$ should agree with Eq.~(\ref{Phires}).
\end{itemize}

In addition, we expect $\Phi(z)\sim z^{3-2/\nu}$ in the limit $z\gg1$, as required in order to ensure that $S(\omega)$ is independent of $\Delta$ when $\omega\gg \Delta$.  Equation (\ref{phistruct}) does not satisfy this requirement.  However, a comparison with experiments or with numerical simulations is bound to be complicated at high frequencies by lattice-scale contributions.  Thus, we choose to focus on the regime in which the frequency is not much greater than $\Delta$.  Then Eq.~(\ref{phistruct}) can serve as a useful prototypical form to fit the data.

Finally, we note that the scaling function $\Phi_\sigma (z)$, which specifies the response function $\chi (\omega)$ via Eq.~(\ref{Sigmascale}),
 has a very similar analytic structure. The only change from $\Phi (z)$ is that 
there is an additional pole at $z =0$. However, the locations of the poles in the lower-half plane are the same, and there is also
a branch cut starting at $z =0$.

\acknowledgements

We thank A.~Auerbach, I.~Bloch, E.~Demler, M.~Endres, L.~Pollet, N.~Prokof'ev, and W.~Zwerger for valuable discussions. 
This research was supported by the National Science Foundation under grant DMR-1103860, 
by a MURI grant from AFOSR, by the Israeli Science Foundation under grant 1338/09, and by a Marie Curie IRG grant.

\appendix
\section{Nonlinear terms in action}
\label{app:lambda}

This appendix will specify the non-linear terms, $\mathcal{S}_1$, in the functional integral over $\lambda$ in Eq.~(\ref{zuj}).
They are given by
\bea
\mathcal{S}_1 &=&  -\frac{i}{6\sqrt{N}} \int_{\bp_1,\bp_2,\bp_3} K_3 (\bp_1,\bp_2, \bp_3) \lambda(\bp_1) \lambda(\bp_2) \lambda (\bp_3) \,\delta \left(\sum_{i=1}^3 \bp_i \right) \nn
&~&~~-  \frac{1}{24N} \int_{\bp_1,\bp_2,\bp_3,\bp_4} K_4 (\bp_1,\bp_2, \bp_3,\bp_4) \lambda(\bp_1) \lambda(\bp_2) \lambda (\bp_3) \lambda(\bp_4) \, \delta \left(\sum_{i=1}^4 \bp_i \right).
\eea
The three-point vertex is given by\cite{abe1}
\bea
K_3 (\bp_1,\bp_2,\bp_3) &=& \int_\bq \frac{1}{(q^2+r) ((\bq+\bp_1)^2 +r)((\bq-\bp_2)^2+r)} \nn
&=& \frac{1}{4 \pi P} \tan^{-1} \frac{P}{\sqrt{r} (p_1^2+p_2^2 + p_3^2 + 8r)} \nn
&\equiv& K_3 (p_1,p_2,p_3),
\eea
where
\beq
P^2 = p_1^2 p_2^2 p_3^2 + r (p_1^2  + p_2^2 + p_3^2)^2 - 2r (p_1^4 + p_2^4 + p_3^4).
\eeq
At the critical point, with $r\rightarrow 0$, we have
\beq
K_3 (p_1,p_2,p_3) =  \frac{1}{8 p_1 p_2 p_3}. \label{l3}
\eeq
At the critical point, we will also need
\bea
K_3 (\bp,-\bp,0) &=& \frac{\Pi (0,0)}{p^2} + \int_\bq \frac{1}{q^4} \left( \frac{1}{(\bq+\bp)^2} - \frac{1}{p^2} \right) \nn
&=& \frac{\Pi (0,0)}{p^2}. \label{l4}
\eea
where we have used the curious fact that
\beq
\int_\bp \frac{1}{p^4} \left( \frac{1}{(\bp + \bq)^2} - \frac{1}{q^2} \right) = 0 \label{curious}
\eeq
only in 3 spacetime dimensions.
Similarly, the 4-point vertex is
\beq
K_4 (\bp_1,\bp_2,\bp_3,\bp_4) = 3\int_\bq \frac{1}{(q^2+r) ((\bq+\bp_1)^2 +r)((\bq+\bp_1+\bp_2)^2+r)((\bq-\bp_4)^2+r)} . 
\eeq
We will not need this expression in general, but only 2 special cases; we have
\bea
K_4 (\bp,\bk,-\bp,-\bk) &=& 3 \int_\bq \frac{1}{(q^2 +r) ((\bq+\bp)^2 + r) ((\bq+\bp+\bk)^2 +r) ((\bq+\bk)^2 + r)} \nn
&=& \frac{3}{2 (\bp \cdot \bk)} \int_\bq \frac{q^2 + r + (\bq+\bp+\bk)^2 + r - (\bq+\bp)^2 -r - (\bq+\bk)^2 - r}{
(q^2 + r) ((\bq+\bp)^2 + r) ((\bq+\bp+\bk)^2 + r) ((\bq+\bk)^2+ r)}\nn
&=& \frac{3}{(\bp \cdot \bk)} \left[ K_3 (k,p,|\bk-\bp|)  - K_3 (k,p,|\bk+\bp|)   \right], \label{l5}
\eea
and from Ref.~\onlinecite{okabe}, at the critical point,
\bea 
K_4 (\bp,-\bp,\bk,-\bk) &=& 3 \int_\bq \frac{1}{q^4 (\bq+\bp)^2  (\bq+\bk)^2 }  \nn
&=& 3 \frac{\Pi (0,0)}{p^2 k^2} +3 \int_\bq \frac{1}{q^4} \left( \frac{1}{(\bq+\bp)^2(\bq+\bk)^2} - \frac{1}{p^2 k^2} \right) \nn
&=& 3 \frac{\Pi (0,0)}{p^2 k^2} + \frac{3 (\bp \cdot \bk)}{8p^3 k^3 |\bp-\bk|}. \label{l6}
\eea

\section{Critical singularities}
\label{app:crit}

This appendix will review the derivation of the quantum critical singularities at $g =g_c$ using the basic expressions in 
Section~\ref{sec:gen}. We will set $u=\infty$ in all expressions here.

Let us first determine the value of the critical coupling $g_c$ to
order $1/N$. The critical point is determined by the
condition $G^{-1} (0) = 0$. So at the critical point $r=r_c$ where, from Eq.~(\ref{gres}),
\bea
r_c &=& -\frac{2}{N} \int_\bq \frac{1}{\Pi (q,0)} \frac{1}{q^2} +
\frac{2}{N} \frac{1}{\Pi (0,0)} \int_{\bp, \bq} \frac{1}{p^4 (\bp + \bq)^2}
\frac{1}{\Pi (q,0)} \nn
&=& \frac{2}{N} \frac{1}{\Pi (0,0)} \int_{q} \frac{1}{\Pi (q,0)}
\int_\bp \frac{1}{p^4} \left( \frac{1}{(\bp + \bq)^2} - \frac{1}{q^2} \right)
\eea
 Note that $\Pi (0,0)$ is infrared divergent, but we assume this is suitably regulated by putting the theory in 
 finite box; we will find that $\Pi (0,0)$ will
eventually cancel out. Using Eq.~(\ref{curious}), we conclude that
\beq
r_c = \mathcal{O} (1/N^2).
\eeq
So from
 Eq.~(\ref{gval}), we obtain
\begin{eqnarray}
\frac{1}{g_c} &=& \int_\bp \frac{1}{p^2} + \mathcal{O} (1/N^2).  \label{gcval}
\end{eqnarray}

Next, we determine the exponent $\eta$. For this we need $G^{-1}
(p)$ at the critical point $r=r_c$. This is given
by
\begin{equation}
G^{-1} (p) = p^2 + \frac{2}{N} \int_\bq \frac{1}{\Pi (q,0)} \left(
\frac{1}{(\bp + \bq)^2} - \frac{1}{q^2} \right) \label{exp1}
\end{equation}
We have
\begin{equation}
G^{-1} (p)) = p^2 \left( 1 + \eta \ln \left(\frac{\Lambda}{p}
\right) \right)
\end{equation}
where
\beq
\eta = \frac{8}{3 \pi^2 N}. \label{eta}
\eeq

For the exponent, $\nu$, assume the coupling is $g$, and define
$r_g$ by
\begin{eqnarray}
\frac{1}{g_c} - \frac{1}{g} &\equiv& \int_\bp \left( \frac{1}{p^2} - \frac{1}{p^2 + r_g} \right) \nonumber \\
&\equiv& \frac{\sqrt{r_g}}{4 \pi}. \label{rgpi}
\end{eqnarray}
Then from Eq.~(\ref{gval}) and (\ref{gcval}) we have
\begin{equation}
r = r_g  + \mathcal{O} (1/N^2). \label{rr}
\end{equation}
Now combining Eq.~(\ref{gres}) and (\ref{rr}), we have the result
for the inverse susceptibility
\begin{eqnarray}
G^{-1} (p) = p^2 + r + F(p,r)
\label{chi}
\end{eqnarray}
where
\begin{eqnarray}
F(p,r) =  \frac{2}{N} \int_\bq \frac{1}{\Pi (q,r)} \left[
\frac{1}{((\bp + \bq)^2 + r)}  + \frac{\Pi^\prime (q,r)}{2\Pi (0, r)}
\right], \label{chi2}
\end{eqnarray}
with
\begin{equation}
\Pi^\prime (p,r) \equiv \frac{\partial \Pi(p,r)}{\partial r} = -2
\int_\bq \frac{1}{(q^2+r)^2((\bp + \bq)^2 + r)} = - \frac{1}{4 \pi \sqrt{r} (p^2 + 4 r)}.
\end{equation}
A useful check is to note that $G^{-1} (0)=0$ as $r \rightarrow 0$.
From (\ref{chi2}) we have
\bea
F(0,r) &=& \frac{12 r}{N\pi} \int_0^\Lambda dq \frac{q^3}{(q^2 + r)(q^2 + 4r) \tan^{-1} (q/(2 \sqrt{r}))} \nn
&\approx & \frac{12}{N \pi^2} r \ln \left( \frac{\Lambda^2}{r} \right)
\eea
From this we obtain the exponent
\beq
\gamma = 2 - \frac{24}{N \pi^2}.
\eeq

Let us now obtain the connected Green's function, $G_{\lambda\lambda} (p)$ to order $1/N$ at the critical
point $r=r_c$. 
We insert (\ref{l3},\ref{l4},\ref{l5},\ref{l6}) into (\ref{l0},\ref{l1}). 
All occurrences of $\Pi (0,0)$ cancel out, and
\bea
G_{\lambda\lambda} (p) &=& 16 p 
- \frac{512}{N} \int_\bk \frac{1}{k |\bk-\bp|} + \frac{256p}{N} \int_\bk 
\frac{1}{(\bp \cdot \bk)} \left[ \frac{1}{|\bk-\bp|}  -  \frac{1}{ |\bk+\bp|} \right] + \frac{512}{N} \int_\bk \frac{(\bp \cdot \bk)}{p k^2 |\bp-\bk|} \nn
&=& 16p - \frac{128}{N \pi^2} \left( 2 \Lambda - p \right) + \frac{256p}{N \pi^2} \ln \left( \frac{\Lambda}{p} \right)
+ \frac{256p}{3 N \pi^2} \left( \ln \left( \frac{\Lambda}{p} \right) + \frac{1}{3} \right)\nn
&=& - \frac{256 \Lambda}{N \pi^2} + 16p \left( 1 + \frac{64}{3 \pi^2 N} \ln \left(\frac{\Lambda}{p}\right)  + \frac{22}{3} \right)
 \label{gll}
\eea
The scaling dimension of $\lambda$ is the same as that of $\phi_\alpha^2$, which is $3-1/\nu$, and so we expect
\beq
G_{\lambda\lambda} (p)  \sim \mbox{constant} + p^{3-2/\nu} 
\eeq
and so (\ref{gll}) is consistent with (\ref{nu}). 

\section{Computation of threshold singularity}
\label{app:thresh}

We describe the evaluation of the $\bk$ integral in Eq.~(\ref{Gth}) at $p = - i (2 \Delta +  \omega) $ as $\omega \rightarrow 0$.

The singularities arising from the integral of $\bk$ in Eq.~(\ref{Gth}) are all at energy $4 \Delta$, and so we only need the explicit
singularities of all the terms as $p \rightarrow - i (2\sqrt{r} +  \omega)$. We list these:
\bea
\Pi (p, r) &=& \frac{ 1 }{16 \pi \sqrt{r}} \ln \left(\frac{4 \sqrt{r}}{-\omega} \right) \nn
\Pi (|\bp + \bk|, r) &=& \frac{ 1 }{16 \pi \sqrt{r}(1 + i k x/\sqrt{r} - k^2/(4 r))^{1/2}} \ln \left(\frac{1+(1 + i k x/\sqrt{r} - k^2/(4 r))^{1/2}}{1-(1 + i k x/\sqrt{r} - k^2/(4 r))^{1/2}} \right) \nn
K_3 (p, k, |\bp+\bk|) &=& \frac{1}{16 \pi \sqrt{r} k (k-2i \sqrt{r} x)} \left[  \ln \left(\frac{4 \sqrt{r}}{-\omega} \right) + 
\ln \left( \frac{k\sqrt{r} (k- 2 i \sqrt{r} x)^2}{(k^2 + 4r) (k - 4 i \sqrt{r} x)} \right) \right] \nn
K_4 (\bp,\bk,-\bp,-\bk) &=&  \frac{3}{16 \pi r k^2 x} \mbox{Im} \left\{  \frac{1}{ (k-2i \sqrt{r} x + (k^2 - 6 i k \sqrt{r}x - 4 r (1+x^2))
\omega/(2 \sqrt{r} (k - 2 i \sqrt{r} x)))} \right. \nn
&~& \quad\quad \quad \times \left. \left[  \ln \left(\frac{4 \sqrt{r}}{-\omega} \right) + 
\ln \left( \frac{k\sqrt{r} (k- 2 i \sqrt{r} x)^2}{(k^2 + 4r) (k - 4 i \sqrt{r} x)} \right) \right] \right\}
\eea
where $x \equiv (\bp \cdot \bk)/(pk)$.

For the second term in Eq.~(\ref{Gth}), we can now pick out the co-efficient of the threshold singularity to be
\bea
&& - \frac{8}{ \pi^2 N} \left[\ln \left( \frac{4 \sqrt{r}}{-\omega} \right) \right]^{-1} \frac{1}{2} \int_{-1}^1 dx \int_0^\infty dk
\ln \left( \frac{k\sqrt{r} (k- 2 i \sqrt{r} x)^2}{(k^2 + 4r) (k - 4 i \sqrt{r} x)} \right) \frac{1}{(k - 2 i \sqrt{r} x)^2 \Pi (|\bk+\bp|,r) \Pi (k,r)} \nn
&& \quad =  \frac{32 \pi \sqrt{r}}{N}\left[\ln \left( \frac{4 \sqrt{r}}{-\omega} \right) \right]^{-1} \times 4.507333003 \label{Gth2}
\eea

For the last term in Eq.~(\ref{Gth}), the term associated with the explicit $\ln (4 \sqrt{r}/(-\omega))$ term in $K_4$ is
\bea
&& \frac{1}{N} \left[\ln \left( \frac{4 \sqrt{r}}{-\omega} \right) \right]^{-1} \frac{1}{2} \int_{-1}^1 dx \int_0^\Lambda dk  \frac{128 k}{x \tan^{-1} ( k/(2 \sqrt{r}))} \nn
&&~~~~\times \mbox{Im} \left\{  \frac{1}{ (k-2i \sqrt{r} x + (k^2 - 6 i k \sqrt{r}x - 4 r (1+x^2))
\omega/(2 \sqrt{r} (k - 2 i \sqrt{r} x)))} \right\} \nn
&& = \frac{32 \pi \sqrt{r}}{N} \left[\ln \left( \frac{4 \sqrt{r}}{-\omega} \right) \right]^{-1} \left\{ 2\ln \left( \frac{4 \sqrt{r}}{-\omega} \right)
+ \frac{16}{\pi^2} \ln\left( \frac{\Lambda}{\sqrt{r}} \right) - 3.44906254 \right\} \label{loglog1}
\eea
The $\ln (\Lambda)$ term above has the same co-efficient as that expected from the corresponding term in Eq.~(\ref{gll}).

The remaining contribution of the last term in Eq.~(\ref{Gth}) is
\bea
&& \frac{1}{N} \left[\ln \left( \frac{4 \sqrt{r}}{-\omega} \right) \right]^{-2} \frac{1}{2} \int_{-1}^1 dx \int_0^\infty dk  \frac{128 k}{x \tan^{-1} ( k/(2 \sqrt{r}))} \ln \left( \frac{k\sqrt{r} (k- 2 i \sqrt{r} x)^2}{(k^2 + 4r) (k - 4 i \sqrt{r} x)} \right)  \nn
&&~~~~\times \mbox{Im} \left\{  \frac{1}{ (k-2i \sqrt{r} x + (k^2 - 6 i k \sqrt{r}x - 4 r (1+x^2))
\omega/(2 \sqrt{r} (k - 2 i \sqrt{r} x)))} \right\} \nn
&& = \frac{32 \pi \sqrt{r}}{N} \left[\ln \left( \frac{4 \sqrt{r}}{-\omega} \right) \right]^{-1} \left\{ - \ln \left( \frac{4 \sqrt{r}}{-\omega} \right)
+ 1.3863 \right\} \label{loglog2}
\eea

Putting the contributions to the last term of Eq.~(\ref{Gth}) in Eqs.~(\ref{loglog1},\ref{loglog2}) together, we have
\bea
&& \frac{4}{3 \Pi^2 (p,r)} \int_\bk \frac{K_4 (\bp,\bk,-\bp,-\bk)}{\Pi (k,r)} = \nn
 &&~~~~~~~~~~ 
 32 \pi \sqrt{r} \left\{1 +  \left[\ln \left( \frac{4 \sqrt{r}}{-\omega} \right) \right]^{-1} \left( \frac{16}{\pi^2} 
 \ln \left( \frac{\Lambda}{\sqrt{r}} \right) - 2.063  \right) \right\} \label{fit}
\eea
Inserting the results in Eqs.~(\ref{Zval}, \ref{Gth2}, \ref{fit}) into Eq.~(\ref{Gth}), we obtain Eq.~(\ref{thresh}). 

\section{Self energies in the Goldstone phase}
\label{app:gold}

We can compute the self-energies associated with the bare connected Green's functions in Eq.~(\ref{bareG}) by a standard $1/N$ expansion
of the partition function in Eq.~(\ref{zhiggs}).
The self-energies to order $1/N$ are 
\bea
\Sigma_{\sigma\sigma} (p) &=&  - \frac{1}{N} \int_\bk G_{\sigma\sigma}^0 (|\bk+\bp|) G_{\lambda \lambda}^0 (k)  - \frac{1}{N} \int_\bk G_{\sigma\lambda}^0 (|\bk+\bp|) G_{\sigma\lambda}^0 (k)   - \frac{1}{N} G_{\sigma\lambda}^0 (0) \int_\bk  G_{\sigma\lambda}^0 (k) \nn
&=&  \frac{256 \sigma_0^2 \Lambda}{\pi^2 N} -  \frac{8(p^2 + 3072 \sigma_0^4)}{3\pi^2 N} \ln \left( \frac{\Lambda}{16\sigma_0^2} \right) + \frac{\sigma_0^4}{N}  F_{\sigma\sigma} (p) \nn
\Sigma_{\lambda\lambda} (p) &=& \frac{1}{2N} \Pi (p,0) - \frac{1}{2N} \int_\bk G_{\sigma\sigma}^0 (|\bk + \bp|) G_{\sigma \sigma}^0 (k)-\frac{1}{2N} \int_\bk \left[K_3 (\bp,\bk,-\bp-\bk) \right]^2 G_{\lambda\lambda}^0 (|\bk + \bp|) G_{\lambda \lambda}^0 (k) 
\nn 
&+& \frac{1}{N} \int_\bk K_3 (\bp,\bk,-\bp-\bk)  G_{\sigma\lambda}^0 (|\bk + \bp|) G_{\sigma \lambda}^0 (k)
+ \frac{1}{N} K_{3} (\bp,-\bp,0) G_{\sigma\lambda}^0 (0) \int_\bk G_{\sigma\lambda}^0 (k) \nn
&+& \frac{1}{6N} \int_\bk K_4 (\bp,\bk,-\bp,-\bk) G_{\lambda\lambda}^0 (k)
+ \frac{1}{3N} \int_\bk K_4 (\bp,-\bp,\bk,-\bk) G_{\lambda\lambda}^0 (k) \nn
&=&   - \frac{ \Lambda}{\pi^2 N p^2} +\frac{4(p + 24 \sigma_0^2)}{3\pi^2 N p^2 } \ln \left( \frac{\Lambda}{16\sigma_0^2} \right) +
\frac{1}{16Np} +  \frac{\sigma_0^2}{Np^2} F_{\lambda\lambda} (p) \nn
\Sigma_{\sigma\lambda} (p) &=& - \frac{1}{N} \int_\bk G_{\sigma\sigma}^0 (|\bk + \bp|) G_{\sigma \lambda}^0 (k) + 
\frac{1}{N}  \int_\bk K_3 (\bp,\bk,-\bp-\bk)  G_{\sigma\lambda}^0 (|\bk + \bp|) G_{\lambda \lambda}^0 (k)\nn
&-& \frac{1}{N} G_{\sigma\sigma}^0 (0) \int_\bk  G_{\sigma\lambda}^0 (k) - \frac{1}{2N} G_{\sigma\lambda}^0 (0) \int_\bk  G_{\sigma\sigma}^0 (k) \nn
&+&\frac{1}{2N} G_{\sigma\lambda}^0 (0) \int_\bk  K_3 (\bk,-\bk,0) G_{\lambda\lambda}^0 (k) 
+ \frac{1}{2N} G_{\sigma\lambda}^0 (0) \int_\bk \frac{1}{k^2} \nn
&=&- \frac{16i \sigma_0 \Lambda}{\pi^2 N p}  + \frac{4 i \sigma_0 (p+128 \sigma_0^2)}{\pi^2 N p} \ln \left( \frac{\Lambda}{16\sigma_0^2} \right) + \frac{i \sigma_0^3}{Np} F_{\sigma\lambda} (p)
\label{selfen}
\eea
Note that all the terms involving $\Pi (0,0)$ in Eq.~(\ref{selfen}) do indeed cancel with each other. In the last line of each expression in Eq.~(\ref{selfen}) we have extracted the ultraviolet divergences explicitly; the functions $F$ are finite and universal dimensionless functions, which depend only on the ratio $p/\sigma_0^2$.  They are given by the following integrals:
\bea
&& F_{\sigma\sigma} (p) = \frac{1}{\sigma_0^4} \int_\bk \left[\frac{k^2 + 256 \sigma_0^4}{\sigma_0^2 (k+ 16 \sigma_0^2) (|\bk + \bp| + 16 \sigma_0^2)} 
- \frac{k^2}{\sigma_0^2 |\bk + \bp| (k + 16 \sigma_0^2)} +  \frac{16}{k+16 \sigma_0^2} \right. \nn
&&~~~~~~~~~~~~~~~~~~~~  \left. - \frac{512 \sigma_0^2}{k^2} + \frac{16(p^2 + 3072 \sigma_0^4)}{3 k^2 (k+16 \sigma_0^2)} \right] \nn
&&= - \int_0^p dk \frac{8 \left(-3 k^2 (p+16 \sigma_0^2)+3 \left(k^2+256 \sigma_0^4\right) k \tanh
   ^{-1}\left(\frac{k}{p+16 \sigma_0^2}\right)+96 k p \sigma_0^2 -p
   \left(p^2+1536 \sigma_0^2\right)\right)}{3 \pi ^2 \sigma_0^4 (k+16 \sigma_0^2) p  } \nn
 &&~~~~ - \int_p^\infty dk \frac{8 \left(3 k \left(k^2+256 \sigma_0^4 \right) \tanh
   ^{-1}\left(\frac{p}{k+16 \sigma_0^2}\right)-p \left(3 k^2-48
   k \sigma_0^2 +p^2+1536 \sigma_0^4 \right)\right)}{3  \pi ^2 \sigma_0^4 (k+16 \sigma_0^2) p } \label{Fdefs1}
\eea
\bea
&& F_{\lambda\lambda} (p) = \frac{p^2}{\sigma_0^2}\int_\bk \left[
\frac{k ( |\bk + \bp| - |\bk - \bp|)}{p ( \bp \cdot \bk ) |\bk + \bp| |\bk - \bp| (k + 16 \sigma_0^2) }
+ \frac{ 2 (\bp \cdot \bk )}{k p^3 |\bk - \bp| (k + 16 \sigma_0^2)} \right. \nn
&&~~\left.
+ \frac{p^2 - 64 k \sigma_0^2 + 64 p \sigma_0^2}{32 k p^2 \sigma_0^2 (k + 16 \sigma_0^2)(|\bk + \bp| + 16 \sigma_0^2)}
- \frac{p + 64  \sigma_0^2}{32 k p \sigma_0^2 |\bk + \bp| (k + 16 \sigma_0^2)} 
  + \frac{2}{k^2 p^2} - \frac{8(p + 24 \sigma_0^2)}{3 p^2 k^2 (k+16 \sigma_0^2)} \right] \nn
&&~~= \int_0^p dk   \Biggr[ \frac{k^2 \log \left(\frac{2 k
   \left(\sqrt{k^2+p^2}+k\right)+p^2}{p^2}\right)}{2\pi^2  \sigma_0^2(k+16\sigma_0^2)
   \sqrt{k^2+p^2}}+\frac{(64 k \sigma_0^2- p^2-64 p \sigma_0^2) \tanh
   ^{-1}\left(\frac{k}{p+16 \sigma_0^2}\right)}{4 \pi ^2 \sigma_0^2 (k+16 \sigma_0^2)
   p} \nn
   &&~~~~~~~~~~~~~~~~~ +\frac{4 k^3-12 k^2 p+12 k p^2-16 p^3-192 p^2 \sigma_0^2}{12 \pi ^2 \sigma_0^2(k+16\sigma_0^2)
   p^2}\Biggr] \nn
&&~~+ \int_p^\infty dk   \Biggr[ \frac{k^2 \log \left(\frac{k^2 + 2 p
   \left(\sqrt{k^2+p^2}+p\right)}{k^2}\right)}{2 \pi ^2 \sigma_0^2(k+16\sigma_0^2)
   \sqrt{k^2+p^2}}+\frac{(64 k \sigma_0^2- p^2-64 p\sigma_0^2) \tanh
   ^{-1}\left(\frac{p}{k+16\sigma_0^2}\right)}{4 \pi ^2\sigma_0^2 (k+16\sigma_0^2)
   p} \nn
   &~&~~~~~~~~~~~~~~~~~ -\frac{p+16\sigma_0^2}{\pi ^2 \sigma_0^2 (k+16\sigma_0^2)
   }\Biggr]  \label{Fdefs2}
\eea
\bea
&& F_{\sigma\lambda} (p) =  \frac{ 1 }{\sigma_0^2} \int_\bk \left[ 
\frac{2k-p}{\sigma_0^2 (k+ 16 \sigma_0^2) (|\bk + \bp| + 16 \sigma_0^2)} 
- \frac{2k-p}{\sigma_0^2 |\bk + \bp| (k + 16 \sigma_0^2)}
- \frac{8 p}{k^2 (k + 16 \sigma_0^2)} \right. \nn
&&~~~~~~~~~~~~~~~~~~~~~~~~~~~~~\left.  + \frac{32}{k^2} - \frac{8 (p + 128 \sigma_0^2)}{k^2 (k+16 \sigma_0^2)} \right] \nn
 &&~~= -\int_0^p dk \frac{8  \left(p (-2 k+p+32 \sigma_0^2)+k (2 k-p) \tanh
   ^{-1}\left(\frac{k}{p+16\sigma_0^2}\right)\right)}{\pi ^2 \sigma_0^2 (k+16\sigma_0^2) p} \nn
   &&~~~~~~ -\int_p^\infty dk \frac{8   \left(p (-2 k+p+32\sigma_0^2)+k (2 k-p) \tanh
   ^{-1}\left(\frac{p}{k+16\sigma_0^2}\right)\right)}{\pi ^2  \sigma_0^2(k+16\sigma_0^2) p}\label{Fdefs3}
 \eea
At small values of $p$ we find
\bea
F_{\sigma\sigma} (p) &=& \frac{4096}{\pi^2} + \frac{40p^2}{9\pi^2\sigma_0^4} +\frac{7
   p^4}{4800 \pi ^2 \sigma_0^8} -\frac{7 p^5}{92160 \pi ^2\sigma_0^10} +\frac{17 p^6}{4816896 \pi
   ^2\sigma_0^{12}} + \mathcal{O}(p^7), \nn
F_{\sigma\lambda} (p) &=& - \frac{256}{\pi^2} - \frac{8 p}{\pi^2\sigma_0^2} - \frac{ p^2}{9 \pi^2\sigma_0^4} 
+\frac{ p^3}{576
   \pi ^2\sigma_0^6}
-\frac{3
   p^4}{5120 \pi ^2\sigma_0^8}
+ \frac{19 p^5}{7372800 \pi ^2\sigma_0^{10}}
+ \mathcal{O} (p^6),    \label{FsmallP}\\
F_{\lambda\lambda} (p) &=& - \frac{16}{\pi^2} - \frac{p}{\pi^2\sigma_0^2} -0.0083038708178 \frac{p^2}{\sigma_0^4} + 0.00077664324643133\frac{p^3}{\sigma_0^6} \nn
&~&~~~+ 0.000022616797187123 \frac{p^4}{\sigma_0^8} - 2.34860390339644 \times 10^{-6}\, \frac{p^5}{\sigma_0^{10}} + 
\mathcal{O} (p^6)  \nn
&~&~~~
+ \left(- \frac{p^3}{384\pi^2\sigma_0^6} + \frac{p^5}{122880\pi^2\sigma_0^{10}} + \mathcal{O} (p^7) \right) \log(p/\sigma_0^2), \label{Fsmall}
\eea
whereas for general values of $p$ these functions can be evaluated numerically. Notice that the $\log (p)$ terms are present
only in $F_{\lambda\lambda}$, and these arise from the integral of the explicitly displayed logarithm in Eq.~(\ref{Fdefs2}).

The imaginary parts of these functions at $p=-16 \sigma_0^2$ are obtained by analytically continuing the contour integration via $k \rightarrow k e^{i \theta}$, $p \rightarrow p e^{i \theta}$ and then taking the limit $\theta \searrow - \pi$:
\bea
\lim_{\theta \searrow - \pi} \mbox{Im} \, F_{\sigma\sigma} (16 \sigma_0^2 e^{i \theta} ) &=&  - \int_0^{-32 \sigma_0^2} \frac{k dk}{4 \pi \sigma_0^6} (k^2+256 \sigma_0^4) \, \mathcal{P} \left( \frac{1}{k+16 \sigma_0^2} \right) - 
 \frac{26624}{3 \pi} = 0 \nn
\lim_{\theta \searrow - \pi} \mbox{Im} \, F_{\sigma\lambda} (16 \sigma_0^2 e^{i \theta}  ) &=&  -\int_0^{-32 \sigma_0^2 } \frac{k dk}{2 \pi \sigma_0^4} (k+8 \sigma_0^2 ) \, \mathcal{P} \left( \frac{1}{k+16 \sigma_0^2 } \right) + \frac{384}{ \pi} = 0 \nn
\lim_{\theta \searrow - \pi} \mbox{Im} \, F_{\lambda\lambda} (-16 \sigma_0^2 e^{i \theta} ) &=&  \int_0^{-32 \sigma_0^2 } \frac{dk}{2 \pi \sigma_0^2} (k+12 \sigma_0^2 ) \, \mathcal{P} \left( \frac{1}{k+16 \sigma_0^2} \right) -\frac{(4 \sqrt{2} \ln(3-2\sqrt{2}))}{\pi} 
\nn &=& 
-\frac{(16 + 4 \sqrt{2} \ln(3-2\sqrt{2}))}{\pi} \label{poleshift}
\eea
The numerical values of these functions evaluated by Mathematica at a series of values of $\theta$ starting at $\theta=0$,
and converging to $\theta=-\pi$, lead to
\bea
F_{\sigma\sigma} (-16 \sigma_0^2 ) &=& 585.93223 + 5 \times 10^{-7} \, i \nn
F_{\sigma\lambda} (-16 \sigma_0^2 ) &=& -19.0308885 + 3 \times 10^{-8} \, i \nn
F_{\lambda\lambda} (-16 \sigma_0^2 ) &=& -1.5589876 - 1.9188981 \,  i ,
\eea
in excellent agreement with the values in Eq.~(\ref{poleshift}). Note that the imaginary values are restricted to $F_{\lambda\lambda}$,
and it is this imaginary contribution which leads to the real part of the pole in Eq.~(\ref{lhppole}).
This imaginary value of $F_{\lambda\lambda}$ is again a consequence of the explicitly displayed logarithm in Eq.~(\ref{Fdefs2}). Thus the $\log (p)$ terms in Eq.~(\ref{Fsmall}) and the oscillatory component of the pole are linked to each other, as is also noted in Section~\ref{sec:conc}.


\begin{thebibliography}{999}

\bibitem{chn} S. Chakravarty, B.I. Halperin, and D.R. Nelson, 
Phys. Rev. Lett. {\bf 60}, 1057 (1988).

\bibitem{fwgf} M.~P.~A.~Fisher, P.~B.~Weichman, G.~Grinstein, and D.~S.~Fisher,
Phys.\ Rev.\ B {\bf 40}, 546 (1989).

\bibitem{csy} A. V. Chubukov, S. Sachdev, and J. Ye, Phys. Rev. B {\bf 49}, 11919 (1994).

\bibitem{ssrelax} S. Sachdev, Phys. Rev. B {\bf 59}, 14054 (1999).

\bibitem{zwerger} W. Zwerger, Phys. Rev. Lett. {\bf 92}, 027203 (2004).

\bibitem{senthil} T.~Senthil, A.~Vishwanath, L.~Balents, S.~Sachdev, and
M.~P.~A.~Fisher, 
Science {\bf 303}, 1490 (2004). 

\bibitem{paa} D.~Podolsky, A.~Auerbach, and D.~P.~Arovas,
Phys. Rev. B {\bf 84}, 174522 (2011).

\bibitem{coldea} R.~Coldea, D.~A.~Tennant, E.~M.~Wheeler,
E.~Wawrzynska, D.~Prabhakaran, M.~Telling, K.~Habicht, P.~Smeibidl, and K.~Kiefer,
Science {\bf 327}, 177 (2010).

\bibitem{sstilt} S. Sachdev, K. Sengupta, and S. M. Girvin, Phys. Rev. B {\bf 66}, 075128 (2002).

\bibitem{simon} J.~Simon, W.~S.~Bakr, R.~Ma, M.~E.~Tai, P.~M.~Preiss, and M.~Greiner, Nature {\bf 472}, 307 (2011).

\bibitem{ruegg} Ch.~Ruegg,  B. Normand, M. Matsumoto, A. Furrer, 
D. F. McMorrow, K. W. Kr\"amer, H.-U. G\"udel, S. N. Gvasaliya, H. Mutka, and M. Boehm,  
Phys. Rev. Lett. {\bf 100}, 205701 (2008).

\bibitem{brown} D.~S.~Chow, P.~Wzietek, D.~Fogliatti, B.~Alavi, D.~J.~Tantillo, C.~A.~Merlic, and S.~E.~Brown, 
Phys. Rev. Lett. {\bf 81}, 3984 (1998).

\bibitem{greiner}  M. Greiner, O. Mandel, T. Esslinger, T. W. H\"ansch, and I. Bloch, 
Nature {\bf 415}, 39 (2002).

\bibitem{spielman} I.~B.~Spielman, W.~D.~Phillips, and J.~V.~Porto, 
Phys. Rev. Lett. {\bf 98}, 080404 (2007).


\bibitem{gemelke} N.~Gemelke, X.~Zhang, C.-L.~Hung, and C. Chin,
Nature {\bf 460}, 995 (2009).

\bibitem{sherson} J.~F. Sherson, C.~Weitenberg, M.~Endres, M.~Cheneau, I.~Bloch I, S. Kuhr,
Nature {\bf 467}, 68 (2010).

\bibitem{bakr} W.~S.~Bakr, A.~Peng, M.~E.~Tai, R.~Ma, J.~Simon, J.~I.~Gillen, S.~F\"olling, L.~Pollet, M.~Greiner,
Science {\bf 329}, 547 (2010).

\bibitem{endresString}  M.~Endres, M.~Cheneau, T.~Fukuhara, C.~Weitenberg, P.~Schau{\ss}, C.~Gross, L.~Mazza, M.~C.~Ba{\~n}uls, L.~Pollet, I.~Bloch, S. Kuhr,
Science {\bf 334}, 200 (2011).
   


\bibitem{endres} M.~Endres, T.~Fukuhara, D.~Pekker, M.~Cheneau, 
P.~Schau{\ss}, C.~Gross, E.~Demler, S.~Kuhr, and I.~Bloch,
Nature {\bf 487}, 454 (2012).
 
\bibitem{wf} K.~G.~Wilson and M.~E.~Fisher, Phys. Rev. Lett. {\bf 28}, 240 (1972).

\bibitem{sushkov} J. Oitmaa, Y. Kulik, and O. P. Sushkov, 
Phys. Rev. B {\bf 85}, 144431 (2012).

\bibitem{ssbook} S.~Sachdev, {\em Quantum Phase Transitions\/}, Second Edition, Cambridge University Press,
Cambridge UK (2011).


\bibitem{lindner}
N.~H.~Lindner and A. Auerbach, Phys. Rev. B {\bf 81}, 054512 (2010).


\bibitem{brezin} E.~Brezin, J.~C.~Le Guillou, and J.~Zinn-Justin, in {\em Phase Transitions and Critical Phenomena}, Vol. 6, edited by C.~Domb and M.~S.~Green (Academic Press, New York, 1976).

\bibitem{pollet} L.~Pollet and N.~V.~Prokof'ev,
Phys. Rev. Lett. {\bf 109}, 010401 (2012).


\bibitem{solvay} S.~Sachdev, in {\em Quantum Theory of Condensed Matter}, Proceedings of the 24th Solvay Conference on Physics, Bertrand Halperin editor (World Scientific, Singapore, 2010), arXiv:0901.4103.

\bibitem{ads2} T.~Faulkner, Hong Liu, J.~McGreevy, and D.~Vegh, 
Phys. Rev. D {\bf 83}, 125002 (2011).

\bibitem{denef} F.~Denef, S.~A.~Hartnoll, and S.~Sachdev, Phys. Rev. D {\bf 80}, 126016 (2009).

\bibitem{hofman} S.~A.~Hartnoll and D.~M.~Hofman, Phys. Rev. B {\bf 81}, 155125 (2010).

\bibitem{abe1} R.~Abe, Prog. Theor. Phys. {\bf 49}, 1877 (1973).

\bibitem{okabe} Y.~Okabe and M.~Oku, Prog. Theor. Phys. {\bf 60}, 1277 (1978).

\end{thebibliography}
\end{document}